%% file: main.tex
\pdfoutput=1

\documentclass[12pt,a4paper]{article}

\usepackage{ifthen} 
\newboolean{pdflatex}
\setboolean{pdflatex}{true} 

\newboolean{articletitles}
\setboolean{articletitles}{true} 

\newboolean{uprightparticles}
\setboolean{uprightparticles}{false} 

\input{preamble}
\usepackage{longtable} 
\usepackage{multirow} 

\begin{document}

\renewcommand{\thefootnote}{\fnsymbol{footnote}}
\setcounter{footnote}{1}

\input{title-LHCb-PAPER}


\renewcommand{\thefootnote}{\arabic{footnote}}
\setcounter{footnote}{0}



\pagestyle{plain} 
\setcounter{page}{1}
\pagenumbering{arabic}


\input{body}

\addcontentsline{toc}{section}{References}
\bibliographystyle{LHCb}
\bibliography{main,LHCb-PAPER,LHCb-CONF,LHCb-DP}

\end{document}

%% file: preamble.tex
\usepackage{microtype}
\usepackage{lineno}  
\usepackage{xspace} 

\usepackage{graphicx}  
\usepackage{color}
\usepackage{booktabs}
\usepackage{colortbl}
\graphicspath{{./}} 

\usepackage{amsmath} 
\usepackage{amssymb}
\usepackage{amsfonts}
\usepackage{upgreek} 

\usepackage{rotating}

\newcommand*\patchAmsMathEnvironmentForLineno[1]{%
\expandafter\let\csname old#1\expandafter\endcsname\csname #1\endcsname
\expandafter\let\csname oldend#1\expandafter\endcsname\csname
end#1\endcsname
 \renewenvironment{#1}%
   {\linenomath\csname old#1\endcsname}%
   {\csname oldend#1\endcsname\endlinenomath}%
}
\newcommand*\patchBothAmsMathEnvironmentsForLineno[1]{%
  \patchAmsMathEnvironmentForLineno{#1}%
  \patchAmsMathEnvironmentForLineno{#1*}%
}
\AtBeginDocument{%
\patchBothAmsMathEnvironmentsForLineno{equation}%
\patchBothAmsMathEnvironmentsForLineno{align}%
\patchBothAmsMathEnvironmentsForLineno{flalign}%
\patchBothAmsMathEnvironmentsForLineno{alignat}%
\patchBothAmsMathEnvironmentsForLineno{gather}%
\patchBothAmsMathEnvironmentsForLineno{multline}%
}

\newcommand{\xx}{\ensuremath{\kern 0.5em }}
\def\pPb {\ensuremath{p\mathrm{Pb}}\xspace}
\def\pA {\ensuremath{p\mathrm{A}}\xspace}

\newcommand{\PromptJpsipA}{\ensuremath{\ 1168\pm\, 15\pm\ \ \, 54\,{\rm \mub}}}
\newcommand{\PromptJpsiAp}{\ensuremath{\ 1293\pm\, 42\pm\ \ \, 75\,{\rm \mub}}}
\newcommand{\JpsiFrombpA}{\ensuremath{ 166.0\pm4.1\pm\ \, 8.2\,{\rm \mub}}}
\newcommand{\JpsiFrombAp}{\ensuremath{ 118.2\pm6.8\pm11.7\,{\rm \mub}}}

\usepackage{hyperref}    
\usepackage[all]{hypcap} 

\input{lhcb-symbols-def} 

\usepackage{cite} 
\usepackage{mciteplus}

%% file: lhcb-symbols-def.tex
\def\lhcb {\mbox{LHCb}\xspace}
\def\ux85 {\mbox{UX85}\xspace}

\ifthenelse{\boolean{uprightparticles}}%
{

 \def\Pmu         {\ensuremath{\upmu}\xspace}

 \def\Ppsi        {\ensuremath{\uppsi}\xspace}

 \def\PDelta      {\ensuremath{\Delta}\xspace}                 
 \def\PXi      {\ensuremath{\Xi}\xspace}                 
 \def\PLambda      {\ensuremath{\Lambda}\xspace}                 
 \def\PSigma      {\ensuremath{\Sigma}\xspace}                 
 \def\POmega      {\ensuremath{\Omega}\xspace}                 
 \def\PUpsilon      {\ensuremath{\Upsilon}\xspace}                 
 
 \def\PB      {\ensuremath{\mathrm{B}}\xspace}                 
                  
 \def\PD      {\ensuremath{\mathrm{D}}\xspace}

 \def\PJ      {\ensuremath{\mathrm{J}}\xspace}                 
 \def\PK      {\ensuremath{\mathrm{K}}\xspace}

 \def\Pi      {\ensuremath{\mathrm{i}}\xspace}

}
{

 \def\Pmu         {\ensuremath{\mu}\xspace}

 \def\Ppsi        {\ensuremath{\psi}\xspace}                 
                  
 \mathchardef\PDelta="7101
 \mathchardef\PXi="7104
 \mathchardef\PLambda="7103
 \mathchardef\PSigma="7106
 \mathchardef\POmega="710A
 \mathchardef\PUpsilon="7107
                  
 \def\PB      {\ensuremath{B}\xspace}                 
                  
 \def\PD      {\ensuremath{D}\xspace}

 \def\PJ      {\ensuremath{J}\xspace}                 
 \def\PK      {\ensuremath{K}\xspace}

 \def\Pi      {\ensuremath{i}\xspace}

}

\def\mmu        {\ensuremath{\Pmu}\xspace}
\def\mup        {\ensuremath{\Pmu^+}\xspace}
\def\mun        {\ensuremath{\Pmu^-}\xspace} 

\def\kaon  {\ensuremath{\PK}\xspace}
  \def\Kbar  {\kern 0.2em\overline{\kern -0.2em \PK}{}\xspace}

\def\Kz    {\ensuremath{\kaon^0}\xspace}
\def\Kzb   {\ensuremath{\Kbar^0}\xspace}
\def\KzKzb {\ensuremath{\Kz \kern -0.16em \Kzb}\xspace}
\def\Kp    {\ensuremath{\kaon^+}\xspace}
\def\Km    {\ensuremath{\kaon^-}\xspace}

\def\KpKm  {\ensuremath{\Kp \kern -0.16em \Km}\xspace}

  \def\Dbar    {\kern 0.2em\overline{\kern -0.2em \PD}{}\xspace}
\def\D       {\ensuremath{\PD}\xspace}

\def\Dz      {\ensuremath{\D^0}\xspace}
\def\Dzb     {\ensuremath{\Dbar^0}\xspace}
\def\DzDzb   {\ensuremath{\Dz {\kern -0.16em \Dzb}}\xspace}
\def\Dp      {\ensuremath{\D^+}\xspace}
\def\Dm      {\ensuremath{\D^-}\xspace}

\def\DpDm    {\ensuremath{\Dp {\kern -0.16em \Dm}}\xspace}

\def\Bbar    {\ensuremath{\kern 0.18em\overline{\kern -0.18em \PB}{}}\xspace}

\def\jpsi     {\ensuremath{{\PJ\mskip -3mu/\mskip -2mu\Ppsi\mskip 2mu}}\xspace}

  \def\Y#1S{\ensuremath{\PUpsilon{(#1S)}}\xspace}

\def\Lbar {\ensuremath{\kern 0.1em\overline{\kern -0.1em\PLambda}}\xspace}

\def\to                 {\ensuremath{\rightarrow}\xspace}

\def\eps   {\ensuremath{\varepsilon}\xspace}

\def\AT#1     {\ensuremath{A_{\mathrm{T}}^{#1}}\xspace}           

\def\C#1      {\ensuremath{\mathcal{C}_{#1}}\xspace}                       
\def\Cp#1     {\ensuremath{\mathcal{C}_{#1}^{'}}\xspace}                    
\def\Ceff#1   {\ensuremath{\mathcal{C}_{#1}^{\mathrm{(eff)}}}\xspace}        
\def\Cpeff#1  {\ensuremath{\mathcal{C}_{#1}^{'\mathrm{(eff)}}}\xspace}       
\def\Ope#1    {\ensuremath{\mathcal{O}_{#1}}\xspace}                       
\def\Opep#1   {\ensuremath{\mathcal{O}_{#1}^{'}}\xspace}                    

\newcommand{\tev}{\ensuremath{\mathrm{\,Te\kern -0.1em V}}\xspace}
\newcommand{\gev}{\ensuremath{\mathrm{\,Ge\kern -0.1em V}}\xspace}
\newcommand{\mev}{\ensuremath{\mathrm{\,Me\kern -0.1em V}}\xspace}
\newcommand{\kev}{\ensuremath{\mathrm{\,ke\kern -0.1em V}}\xspace}
\newcommand{\ev}{\ensuremath{\mathrm{\,e\kern -0.1em V}}\xspace}
\newcommand{\gevc}{\ensuremath{{\mathrm{\,Ge\kern -0.1em V\!/}c}}\xspace}
\newcommand{\mevc}{\ensuremath{{\mathrm{\,Me\kern -0.1em V\!/}c}}\xspace}
\newcommand{\gevcc}{\ensuremath{{\mathrm{\,Ge\kern -0.1em V\!/}c^2}}\xspace}
\newcommand{\gevgevcccc}{\ensuremath{{\mathrm{\,Ge\kern -0.1em V^2\!/}c^4}}\xspace}
\newcommand{\mevcc}{\ensuremath{{\mathrm{\,Me\kern -0.1em V\!/}c^2}}\xspace}

\def\mm   {\ensuremath{\rm \,mm}\xspace}

\def\mum  {\ensuremath{\,\upmu\rm m}\xspace}

\def\fm   {\ensuremath{\rm \,fm}\xspace}

\def\mub{\ensuremath{{\rm \,\upmu b}}\xspace}

\def\deriv {\ensuremath{\mathrm{d}}}

\def\gsim{{~\raise.15em\hbox{$>$}\kern-.85em
          \lower.35em\hbox{$\sim$}~}\xspace}
\def\lsim{{~\raise.15em\hbox{$<$}\kern-.85em
          \lower.35em\hbox{$\sim$}~}\xspace}

\def\sPlot{\mbox{\em sPlot}\xspace}

\def\pt         {\mbox{$p_{\rm T}$}\xspace}

\def\dllmupi    {\ensuremath{\mathrm{DLL}_{\mmu\pi}}\xspace}

\def\evtgen     {\mbox{\textsc{EvtGen}}\xspace}
\def\pythia     {\mbox{\textsc{Pythia}}\xspace}

\def\geant      {\mbox{\textsc{Geant4}}\xspace}

\def\photos     {\mbox{\textsc{Photos}}\xspace}

\def\tell1  {TELL1\xspace}
\def\ukl1   {UKL1\xspace}

\newcommand{\ie}{\mbox{\itshape i.e.}\xspace}

%% file: title-LHCb-PAPER.tex

\begin{titlepage}
\pagenumbering{roman}

\vspace*{-1.5cm}
\centerline{\large EUROPEAN ORGANIZATION FOR NUCLEAR RESEARCH (CERN)}
\vspace*{1.5cm}
\hspace*{-0.5cm}
\begin{tabular*}{\linewidth}{lc@{\extracolsep{\fill}}r}
\ifthenelse{\boolean{pdflatex}}
{\vspace*{-2.7cm}\mbox{\!\!\!\includegraphics[width=.14\textwidth]{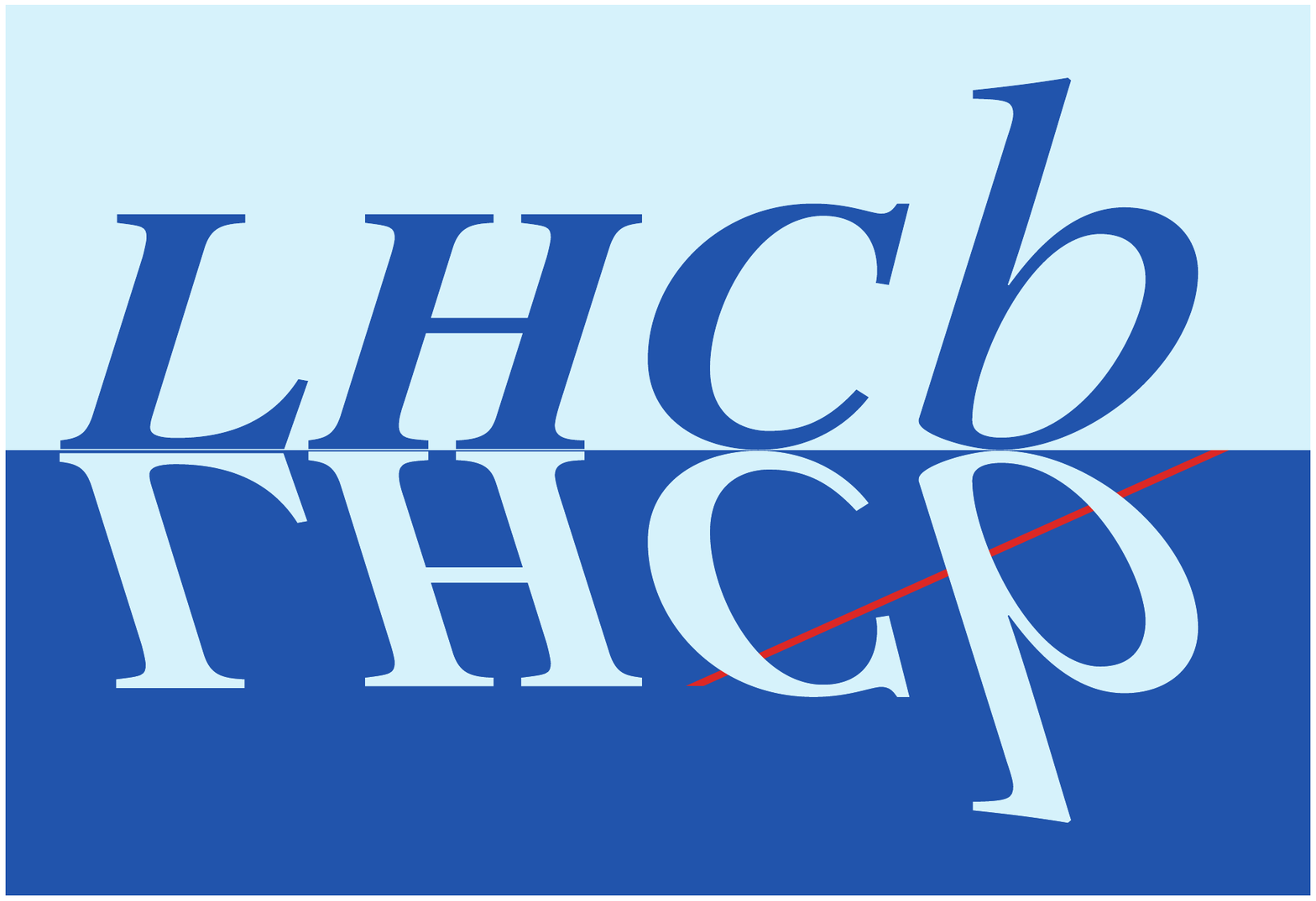}} & &}%
{\vspace*{-1.2cm}\mbox{\!\!\!\includegraphics[width=.12\textwidth]{lhcb-logo.eps}} & &}%
\\
 & & CERN-PH-EP-2013-156 \\  
 & & LHCb-PAPER-2013-052 \\  
 & & 30 August, 2013 \\ 
\end{tabular*}

\vspace*{1.5cm}

{\bf\boldmath\huge
\begin{center}
      Study of $\jpsi$ production and cold nuclear matter effects
      in $p\mathrm{Pb}$ collisions 
      at $\sqrt{s_{\mbox{\small{\it NN}}}}=5\mathrm{\,Te\kern -0.1em V}$
\end{center}
}

\vspace*{1.5cm}

\begin{center}
The LHCb collaboration\footnote{Authors are listed on the following pages.}
\end{center}

\vspace{\fill}

\begin{abstract}
  \noindent
  The production of $\jpsi$ mesons with rapidity $1.5<y<4.0$ or $-5.0<y<-2.5$
  and transverse momentum $\pt<14\gevc$ is studied with the {\mbox{LHCb}\xspace}\ detector 
  in proton-lead collisions at a nucleon-nucleon centre-of-mass energy  
  $\sqrt{s_{\mbox{\tiny{\it NN}}}}=5\mathrm{\,Te\kern -0.1em V}$. 
  The \jpsi mesons are reconstructed using the dimuon decay mode.
  The analysis is based on a data sample corresponding to an integrated luminosity of 
  about $1.6~\mathrm{nb}^{-1}$.
  For the first time the nuclear modification factor and forward-backward production ratio
  are determined separately for prompt $\jpsi$ mesons and $\jpsi$ from $b$-hadron decays.
  Clear suppression of prompt $\jpsi$ production with respect to proton-proton collisions 
  at large rapidity is observed, while the production of $\jpsi$ from $b$-hadron decays is less suppressed.
  These results show good agreement with available theoretical predictions.
  The measurement shows that cold nuclear matter effects are important for 
  interpretations of the related quark-gluon plasma signatures in heavy-ion collisions.
\end{abstract}

\vspace*{2.0cm}

\begin{center}
  Submitted to JHEP 
\end{center}

\vspace{\fill}

{\footnotesize 
\centerline{\copyright~CERN on behalf of the \lhcb collaboration, license \href{http://creativecommons.org/licenses/by/3.0/}{CC-BY-3.0}.}}
\vspace*{2mm}

\end{titlepage}


\newpage
\setcounter{page}{2}
\mbox{~}
\newpage

\input{LHCb_authorlist.tex}

\cleardoublepage

%% file: LHCb_authorlist.tex
\centerline{\large\bf LHCb collaboration}
\begin{flushleft}
\small
R.~Aaij$^{40}$, 
B.~Adeva$^{36}$, 
M.~Adinolfi$^{45}$, 
C.~Adrover$^{6}$, 
A.~Affolder$^{51}$, 
Z.~Ajaltouni$^{5}$, 
J.~Albrecht$^{9}$, 
F.~Alessio$^{37}$, 
M.~Alexander$^{50}$, 
S.~Ali$^{40}$
% ...

{\footnotesize \it
% ...
$ ^{5}$Clermont Universit\'{e}, Universit\'{e} Blaise Pascal, CNRS/IN2P3, LPC, Clermont-Ferrand, France\\
$ ^{6}$CPPM, Aix-Marseille Universit\'{e}, CNRS/IN2P3, Marseille, France\\
% ...
$ ^{9}$Fakult\"{a}t Physik, Technische Universit\"{a}t Dortmund, Dortmund, Germany\\
% ...
$ ^{36}$Universidad de Santiago de Compostela, Santiago de Compostela, Spain\\
$ ^{37}$European Organization for Nuclear Research (CERN), Geneva, Switzerland\\
% ...
$ ^{40}$Nikhef National Institute for Subatomic Physics, Amsterdam, The Netherlands\\
% ...
$ ^{45}$H.H. Wills Physics Laboratory, University of Bristol, Bristol, United Kingdom\\
% ...
$ ^{50}$School of Physics and Astronomy, University of Glasgow, Glasgow, United Kingdom\\
$ ^{51}$Oliver Lodge Laboratory, University of Liverpool, Liverpool, United Kingdom\\
% ...
}
\end{flushleft}

%% file: body.tex
\section{Introduction}
\label{sec:Introduction}
\noindent 
The suppression of heavy quarkonia production
with respect to proton-proton ($pp$) collisions~\cite{Matsui:1986dk}
is one of the most distinctive signatures of the formation of quark-gluon plasma, 
a hot nuclear medium 
created in ultrarelativistic heavy-ion collisions.
However, the suppression of heavy quarkonia and light hadron production 
with respect to $pp$ collisions
can also take place in proton-nucleus ($p$A) collisions, 
where a quark-gluon plasma is not expected to be created and only cold nuclear matter effects,
such as nuclear absorption, parton shadowing and parton energy loss in initial and 
final states occur~\cite{Ferreiro:2013pua,Albacete:2013ei,Arleo:2012rs,Arleo:2013zua,Arleo:2012hn,Chirilli:2012jd,Chirilli:2012sk}.
The study of $p$A collisions is important to disentangle the effects of 
quark-gluon plasma from cold nuclear matter,
and to provide essential input to the understanding of nucleus-nucleus collisions. 
Nuclear effects are usually characterised by the nuclear modification factor,
defined as the production cross-section of a given particle in $p$A collisions 
divided by that in $pp$ collisions
and the number of colliding nucleons in the nucleus (given by the atomic number $A$),
\begin{equation}
\label{eq:AttenuationFactor}
 R_{\pA}(y,p_\mathrm{T},\sqrt{s_{\mbox{\tiny{\it NN}}}})
    \equiv\frac{1}{A}
    \frac{\deriv^2\sigma_{\pA}(y,\pt,\sqrt{s_{\mbox{\tiny{\it NN}}}})/\deriv y\deriv\pt}
         {\deriv^2\sigma_{pp}(y,\pt,\sqrt{s_{\mbox{\tiny{\it NN}}}})/\deriv y\deriv\pt},
\end{equation}
where 
$y$ is the rapidity of the particle in the nucleon-nucleon centre-of-mass frame, 
$p_\mathrm{T}$ is the transverse momentum of the particle,
and $\sqrt{s_{\mbox{\tiny{\it NN}}}}$\/ is the nucleon-nucleon centre-of-mass energy.
The suppression of heavy quarkonia and light hadron production
with respect to $pp$ collisions
at large rapidity has been observed 
in $p$A collisions~\cite{Leitch:1999ea,Abt:2008ya}
and in deuteron-gold collisions~\cite{Arsene:2004ux,Adler:2004eh,Adare:2010fn}, 
but has not been studied in proton-lead ($p$Pb) collisions 
at the $\mathrm{\,Te\kern -0.1em V}$ scale.
Previous experiments~\cite{Leitch:1999ea,Arsene:2004ux,Adler:2004eh,Abt:2008ya,Adare:2010fn} 
have also shown evidence that the production cross-section of $\jpsi$ mesons 
or light hadrons
in the forward region (positive rapidity) of $p$A or deuteron-gold collisions
differs from that in the backward region (negative rapidity),
where ``forward'' and ``backward'' are defined relative to the direction 
of the proton or deuteron beam. 
Measurements of the nuclear modification factor $R_{\pPb}$ 
and the forward-backward production ratio 
\begin{equation}
\label{eq:RFB}
R_{\mbox{\tiny{FB}}}(y,\pt,\sqrt{s_{\mbox{\tiny{\it NN}}}})\equiv
\frac{\deriv^2\sigma_{\pPb}(+|y|,\pt,\sqrt{s_{\mbox{\tiny{\it NN}}}})/\deriv y\deriv\pt}
     {\deriv^2\sigma_{\pPb}(-|y|,\pt,\sqrt{s_{\mbox{\tiny{\it NN}}}})/\deriv y\deriv\pt} 
\end{equation}
are sensitive to
cold nuclear matter effects. 
The advantage of measuring the ratio $R_{\mbox{\tiny{FB}}}$ 
is that it does not rely on the knowledge of the  
$\jpsi$ production cross-section in $pp$ collisions.
Another advantage is that part of experimental systematic uncertainties 
and of the theoretical scale uncertainties 
cancel out in the ratio.

The asymmetric layout of the {\mbox{LHCb}\xspace} 
experiment~\cite{Alves:2008zz}, covering the \mbox{pseudorapidity} range $2<\eta <5$,
allows for a measurement of $R_{\pPb}$\/ for both the forward 
and backward regions, taking advantage of the inversion of 
the proton and lead beams during the $\pPb$ data-taking period in 2013. 
The energy of the proton beam is $4\mathrm{\,Te\kern -0.1em V}$,
while that of the lead beam is $1.58\mathrm{\,Te\kern -0.1em V}$ per nucleon,
resulting in a centre-of-mass energy of the nucleon-nucleon system of $5.02\tev$,
approximated as $\sqrt{s_{\mbox{\tiny{\it NN}}}}=5\mathrm{\,Te\kern -0.1em V}$
due to the uncertainty of the beam energy.
Since the energy per nucleon in the proton beam is significantly larger
than that in the lead beam,
the nucleon-nucleon centre-of-mass system has a rapidity 
in the laboratory frame of $+0.465$ ($-0.465$) for $\pPb$ 
forward (backward) collisions.
This results in a shift of the rapidity coverage in the nucleon-nucleon
centre-of-mass system, ranging from about $1.5$ to $4.0$ for forward $\pPb$ 
collisions and from $-5.0$ to $-2.5$ for backward $\pPb$ collisions. 
The excellent vertexing capability of LHCb allows a separation of
prompt $\jpsi$ mesons
and $\jpsi$ mesons from $b$-hadron decays 
(abbreviated as ``$\jpsi$ from $b$" in the following).
The sum of these two components is referred to as inclusive $\jpsi$ mesons.

In this paper, the differential production cross-sections of prompt $\jpsi$ mesons
and $\jpsi$ from $b$, as functions of $y$ and $\pt$, 
are measured for the first time in $\pPb$
collisions at $\sqrt{s_{\mbox{\tiny{\it NN}}}}=5\mathrm{\,Te\kern -0.1em V}$.
Measurements of $R_{\pPb}$ and $R_{\mbox{\tiny{FB}}}$, 
for both prompt $\jpsi$ mesons and $\jpsi$ from $b$, 
are presented.
For the ease of the comparison with other experiments, 
results for inclusive $\jpsi$ mesons are also given.

\vspace{-0.1cm}
\section{Detector and data set}
\label{sec:Detector}
The \mbox{LHCb}\xspace detector~\cite{Alves:2008zz} is a single-arm forward
spectrometer 
designed for the study of particles containing $b$ or $c$ 
quarks. The detector includes a high precision tracking system
consisting of a silicon-strip vertex detector ({VELO\xspace})
surrounding the $pp$\/
interaction region, a large-area silicon-strip detector located
upstream of a dipole magnet with a bending power of about
$4{\rm\,Tm}$, and three stations of silicon-strip detectors and straw
drift tubes placed downstream. 
The VELO has the unique feature of being located very close to the beam line 
(about $8\,\mm$).  
This allows excellent resolutions in reconstructing 
the position of the collision point, \ie, the primary vertex,
and the vertex of the hadron decay, \ie, the secondary vertex.
For primary (secondary) vertices, the resolution in the plane
transverse to the beam is $\sigma_{x,y}\approx10~(20)\mum$, 
and that along the beam is $\sigma_{z}\approx50~(200)\mum$.
The combined tracking system has
a momentum resolution $\Delta p/p$\/ that varies from 0.4\% 
at $5\mathrm{\,Ge\kern -0.1em V\!/}c$ to
0.6\% at $100\mathrm{\,Ge\kern -0.1em V\!/}c$, and an impact parameter resolution 
of $20\,\upmu\mathrm{m}$ for
tracks with large transverse momentum. Charged hadrons are identified
using two ring-imaging Cherenkov detectors~\cite{LHCb-DP-2012-003}. 
Photon, electron and hadron candidates are identified by a calorimeter system 
consisting of scintillating-pad and preshower detectors, an electromagnetic
calorimeter and a hadronic calorimeter. Muons are identified by a
system composed of alternating layers of iron and multiwire
proportional chambers~\cite{LHCb-DP-2012-002}.
The trigger~\cite{LHCb-DP-2012-004} consists of a
hardware stage, based on information from the calorimeter and muon
systems, followed by a software stage which applies a full event
reconstruction.

This analysis is based on a data sample acquired during the $\pPb$
run in early 2013, corresponding to an integrated luminosity of 
$1.1~\mathrm{nb}^{-1}$ ($0.5~\mathrm{nb}^{-1}$)
for forward (backward) collisions. 
The instantaneous luminosity was 
around $5\times10^{27}\,\mathrm{cm}^{-2}\mathrm{s}^{-1}$,
five orders of magnitude below the typical LHCb luminosity for $pp$\/ collisions. 

The hardware trigger during this period was simply an interaction trigger,
which rejects empty events.
The software trigger requires one well-reconstructed track with hits
in the muon system and a $p_{\mathrm{T}}$ 
greater than $600\mathrm{\,Me\kern -0.1em V\!/}c$.

Simulated samples based on $pp$\/ collisions at $8\mathrm{\,Te\kern -0.1em V}$ 
are reweighted to reproduce the experimental data at $5\mathrm{\,Te\kern -0.1em V}$,
and are used to determine acceptance and reconstruction efficiencies,
where the effect of the asymmetric beam energies in $\pPb$ collisions has been properly taken into account.
In the simulation, $pp$ collisions are generated using
\pythia~6.4~\cite{Sjostrand:2006za} with a specific \lhcb
configuration~\cite{LHCb-PROC-2010-056}. Hadron decays 
are described by \evtgen~\cite{Lange:2001uf}, where final state
radiation is generated using \photos~\cite{Golonka:2005pn}. The
interactions of the generated particles with the detector and its
response are implemented using the \geant
toolkit~\cite{Allison:2006ve, *Agostinelli:2002hh} as described in
Ref.~\cite{LHCb-PROC-2011-006}.

\section{Event selection and cross-section determination}
\label{sec:JpsiAnalysis}
The $\jpsi$\ production cross-section measurement 
follows the approach described in
Refs.~\cite{LHCb-PAPER-2011-003,LHCb-PAPER-2012-039,LHCb-PAPER-2013-016}.
The $\jpsi$ candidates are reconstructed and selected using 
dimuon final states in the events with at least one primary vertex,
which consists of no less than five tracks.
Reconstructed $\jpsi\to\mu^+\mu^-$ candidates are selected 
from pairs of oppositely charged particles 
with transverse momentum $p_\mathrm{T}>0.7\mathrm{\,Ge\kern -0.1em V\!/}c$,
which are identified as muons by the muon detector
and have a track fit $\chi^2$ per number of degree of freedom less than $3$.
To suppress combinatorial background, 
the difference between the logarithms of the likelihoods for 
the muon and the pion hypotheses $\dllmupi$~\cite{LHCb-DP-2012-002,LHCb-DP-2013-001} 
is required to be greater than $1.0~(3.5)$ for the forward (backward) sample.
The two muons are required to originate from a common vertex with 
a $\chi^2$-probability larger than $0.5\%$. 
Candidates are kept if the reconstructed invariant mass is 
in the range $2990<m_{\mu\mu}<3210\mathrm{\,Me\kern -0.1em V\!/}c^2$,
which is within about $\pm110\mathrm{\,Me\kern -0.1em V\!/}c^2$
of the known $\jpsi$ mass~\cite{PDG2012}.

The double differential cross-section for $\jpsi$\/ production 
in a given $(p_{\mathrm{T}},y)$ bin is defined as 
\begin{equation}
 \label{eq:JpsiDoubleDifferential}
  \frac{\mathrm{d}^2\sigma}{\mathrm{d} p_{\mathrm{T}}\mathrm{d}{}y}
 =\frac{N^{\mathrm{cor}}(\jpsi\to\mu^+\mu^-)}
       {\mathcal{L}\times\mathcal{B}(\jpsi\to\mu^+\mu^-)\times\Delta p_{\mathrm{T}}\times\Delta{}y},
\end{equation}
where $N^\mathrm{cor}(\jpsi\to\mu^+\mu^-)$\/ is the efficiency-corrected number 
of observed $\jpsi\to\mu^+\mu^-$\/ signal candidates in the given bin, 
$\mathcal{L}$ is the integrated luminosity,
$\mathcal{B}(\jpsi\to\mu^+\mu^-)=(5.93\pm0.06)\%$~\cite{PDG2012} 
is the branching fraction of the $\jpsi\to\mu^+\mu^-$ decay, 
and $\Delta p_{\mathrm{T}}$\/ and $\Delta{}y$\/ 
the widths of the $(p_\mathrm{T},\, y)$ bin.

The numbers of prompt $\jpsi$ mesons and $\jpsi$ from $b$ 
in bins of the kinematic variables $y$ and $p_\mathrm{T}$
are obtained by performing combined extended maximum likelihood fits 
to the unbinned distributions of dimuon mass 
and pseudo proper time $t_z$\/ in each kinematic bin. 
The pseudo proper time of the $\jpsi$\ meson is defined as
\begin{equation}
 \label{JpsiPseudoProperTime}
 t_z=\frac{(z_\jpsi-z_{\mathrm{PV}})\times{}M_\jpsi}{p_z},
\end{equation}
where $z_\jpsi$\/ is the $z$\/ position of the \jpsi\ decay vertex,
$z_{\mathrm{PV}}$\/ that of the primary vertex,
$p_z$\/ is the $z$ component of the measured $\jpsi$\ momentum,
and $M_{\jpsi}$\/ is the known $\jpsi$\ mass~\cite{PDG2012}.

The signal dimuon invariant mass distribution in each $p_\mathrm{T}$ and $y$ bin is 
modelled with a Crystal Ball function~\cite{Skwarnicki:1986xj},
and the combinatorial background with an exponential function.
The $t_z$ signal distribution is described by the sum of a $\delta$-function 
at $t_z=0$ for prompt $\jpsi$ production
and an exponential decay function for $\jpsi$ from $b$,
both convolved with a double-Gaussian resolution function whose parameters are free in the fit.
The $t_z$ distribution of background in each kinematic bin is independently modelled 
with an empirical function based on the $t_z$ distribution observed in background events
obtained using the {\mbox{\em sPlot}}\ technique~\cite{Pivk:2004ty}.
All the parameters of the $t_z$ background distribution are fixed 
in the final combined fits to the distributions of invariant mass and pseudo proper time.
The total fit function is the sum of the products of the mass and $t_z$ fit functions
for the signal and background components.

\begin{figure}[tb]
  \begin{center}
   \includegraphics[width=0.49 \textwidth]{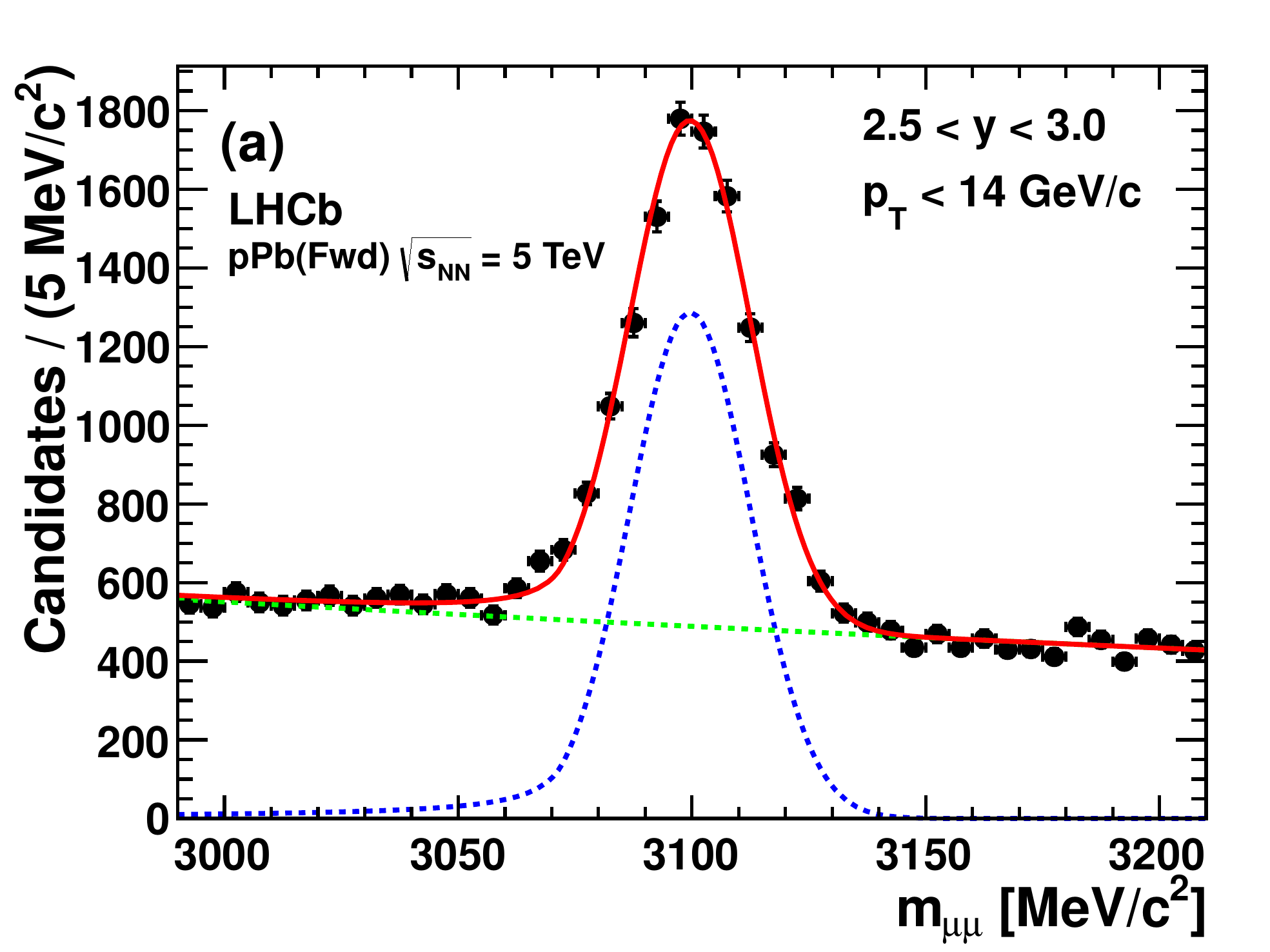}
   \includegraphics[width=0.49 \textwidth]{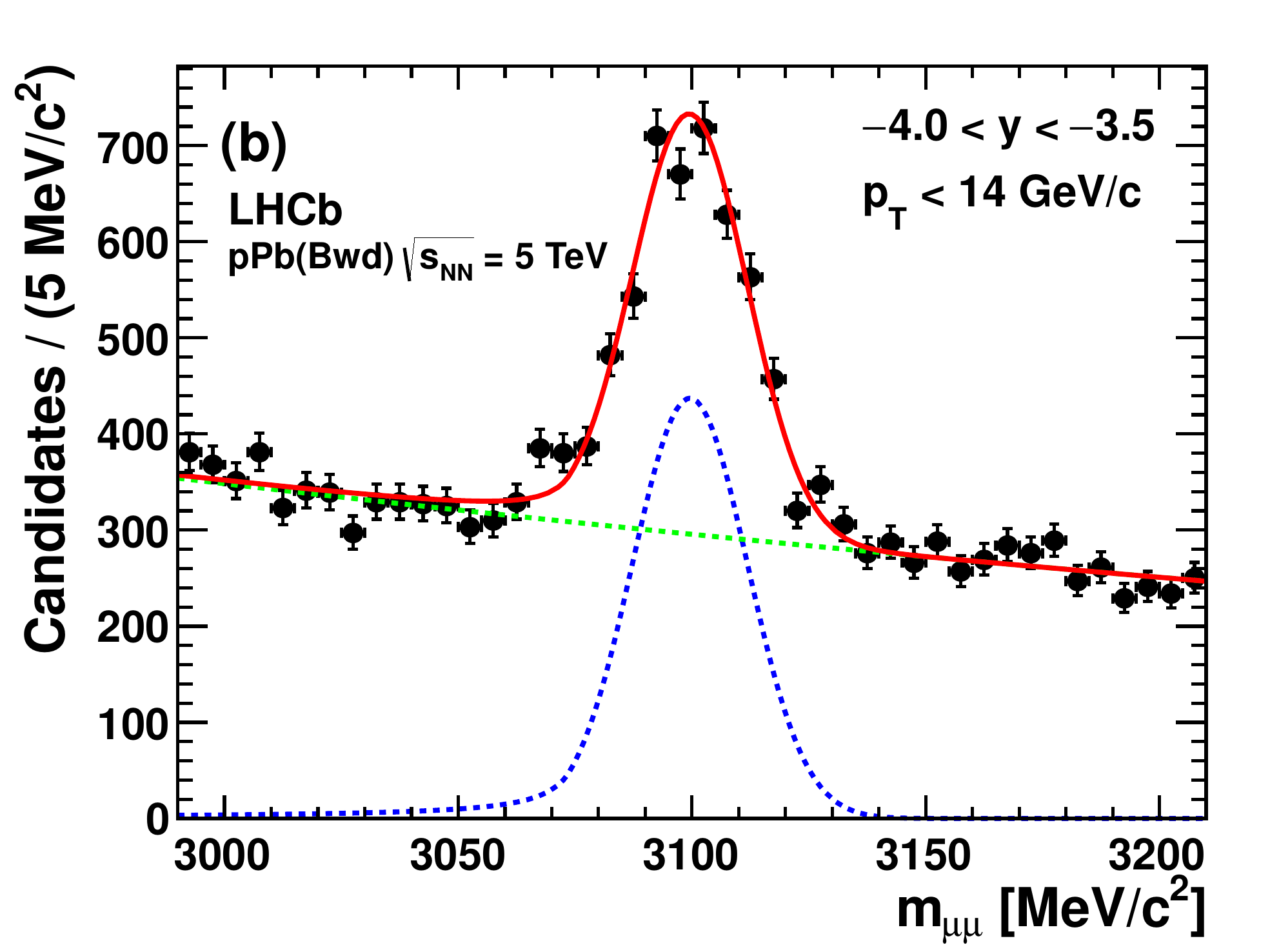}
   \includegraphics[width=0.49 \textwidth]{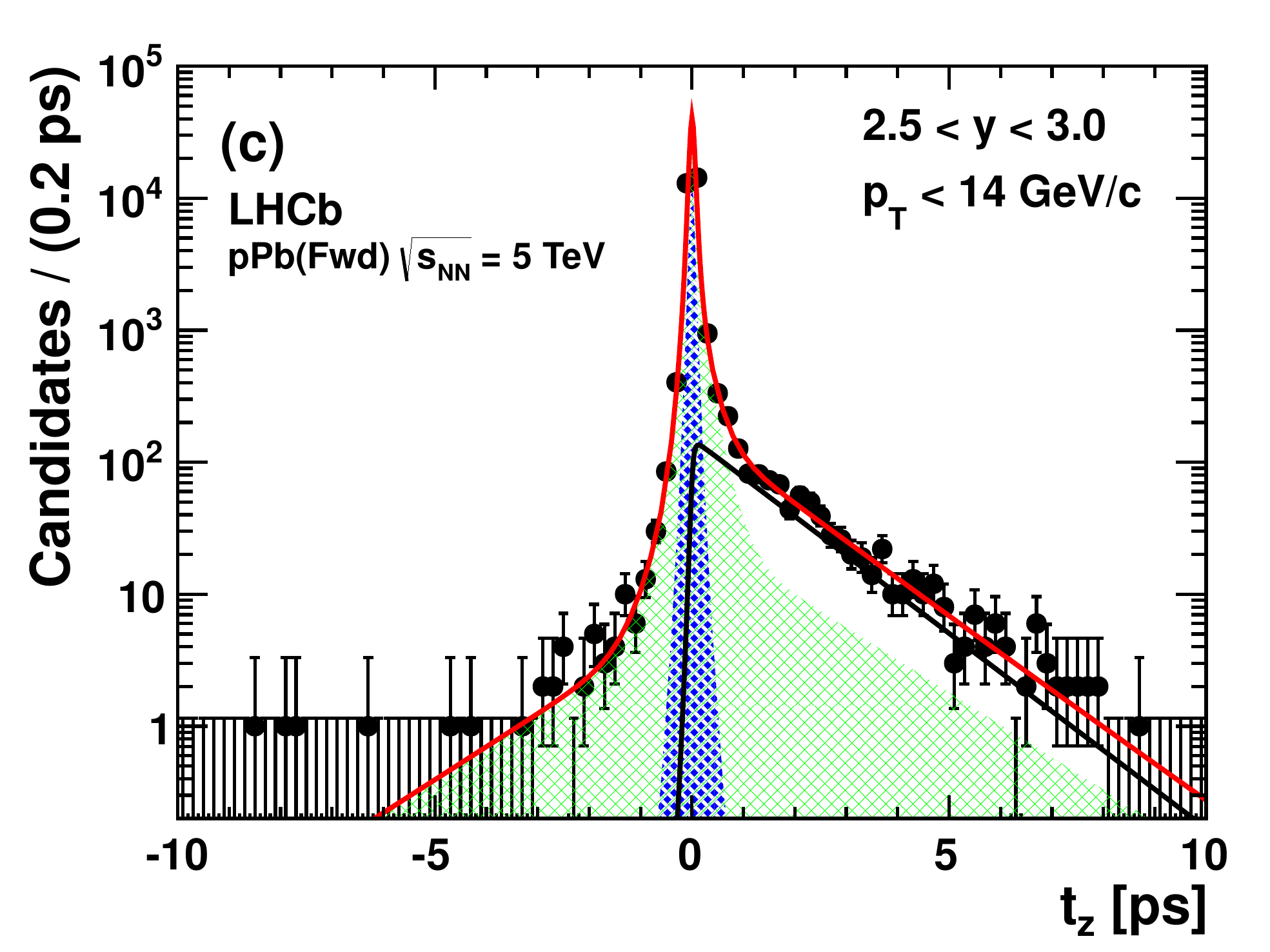}
   \includegraphics[width=0.49 \textwidth]{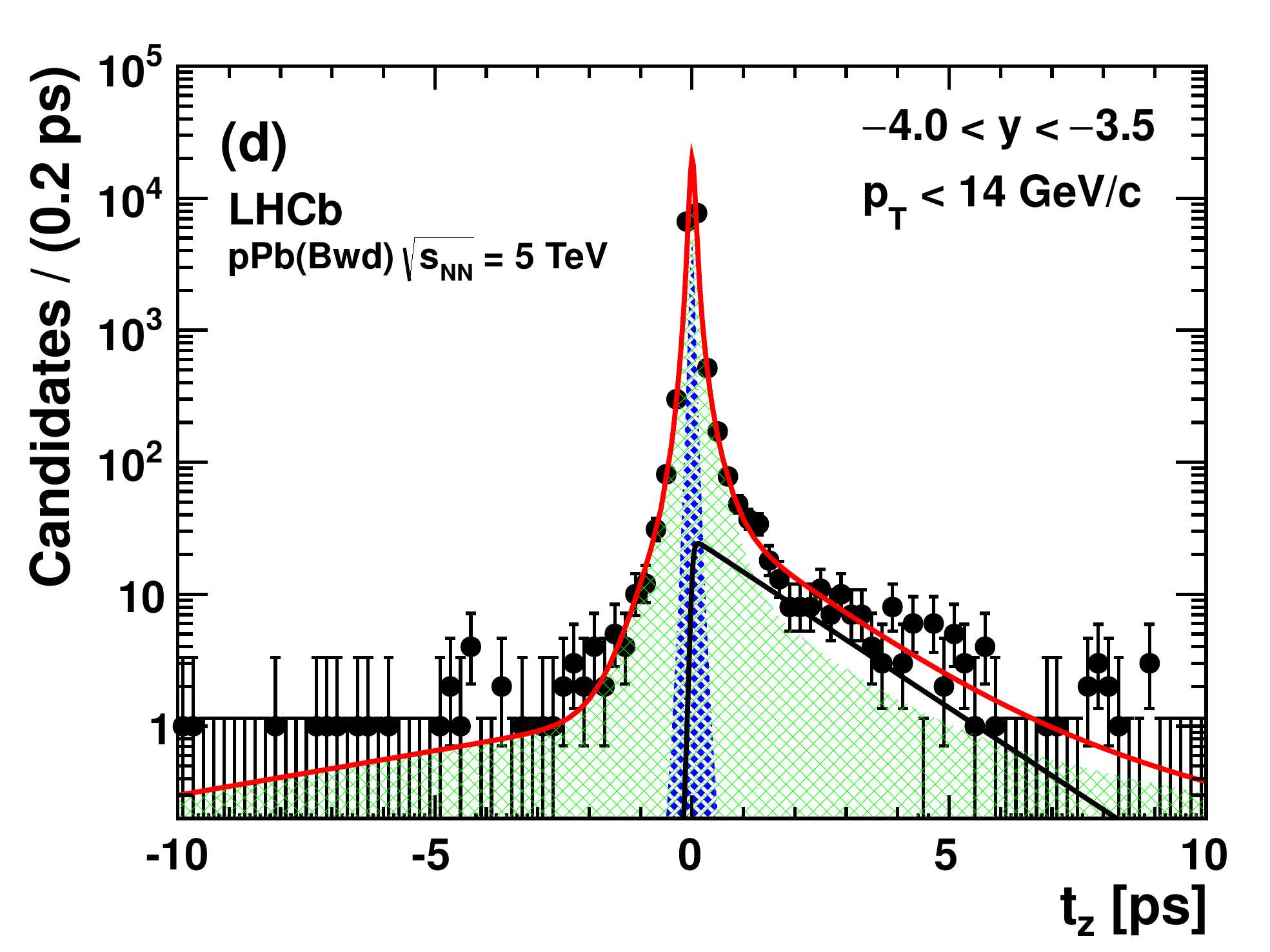}
   \vspace*{-1.0cm}
  \end{center}
 \caption{\small 
     Projections of the combined fit on (a, b) dimuon invariant mass 
     and (c, d) $t_z$ in two representative bins 
     in the (a, c) 
     forward 
     and (b, d) 
     backward samples.
     For the mass projections the (red solid curve) total fitted function is shown 
     together with the (blue dotted curve) $\jpsi$ signal 
     and (green dotted curve) background contributions.
     For the $t_z$ projections the total fitted function is indicated 
     by the solid red curve, the background by the green hatched area, 
     the prompt signal by the blue area and $\jpsi$ from $b$
     by the solid black curve.
 }
 \label{fig:tzFit}
\end{figure}

Figure \ref{fig:tzFit} shows projections of the fit to the dimuon invariant mass 
and $t_z$ distributions, 
for two representative bins of $y$ in the forward and backward regions. 
Higher combinatorial background in the backward region is seen
due to its larger multiplicity.
The dimuon invariant mass resolution is about $15\mevcc$ 
for both the forward and backward samples, consistent with the mass resolution 
measured in $pp$ 
collisions~\cite{LHCb-PAPER-2011-003,LHCb-PAPER-2012-039,LHCb-PAPER-2013-016} 
and in simulation.
The total signal yield for prompt $\jpsi$ mesons in the forward (backward) sample
is $25\,280\pm240$ ($8\,830\pm160$), and the total signal yield for $\jpsi$ from $b$ 
in the forward (backward) sample is $3\,720\pm80$ ($890\pm40$),
where the uncertainty is statistical.
Based on the fit results for prompt $\jpsi$\ mesons
and $\jpsi$\ from $b$, a signal weight factor $w_i$\/ for the $i$th candidate
is obtained with the {\mbox{\em sPlot}}\ technique, 
using the dimuon invariant mass and $t_z$ as discriminating variables. 
The sum of $w_i/\eps_i$\/ over all events in a given bin leads to the 
efficiency-corrected signal yield $N^\mathrm{cor}$ in that bin, 
where the efficiency $\eps_i$ depends on $p_\mathrm{T}$ and $y$ 
and includes the geometric acceptance, reconstruction, muon identification, 
and trigger efficiencies.

The acceptance and reconstruction efficiencies are estimated from simulated samples, 
assuming production of unpolarised $\jpsi$ mesons.
The efficiency of the \dllmupi selection is obtained 
by a data-driven tag-and-probe approach~\cite{Jaeger:1402577}.
The trigger efficiency is obtained from data using
a sample of $\jpsi$\ decays unbiased by the trigger decision~\cite{LHCb-DP-2012-004}.

\begin{figure}[tb]
\begin{center}
\includegraphics[width=0.49 \textwidth]{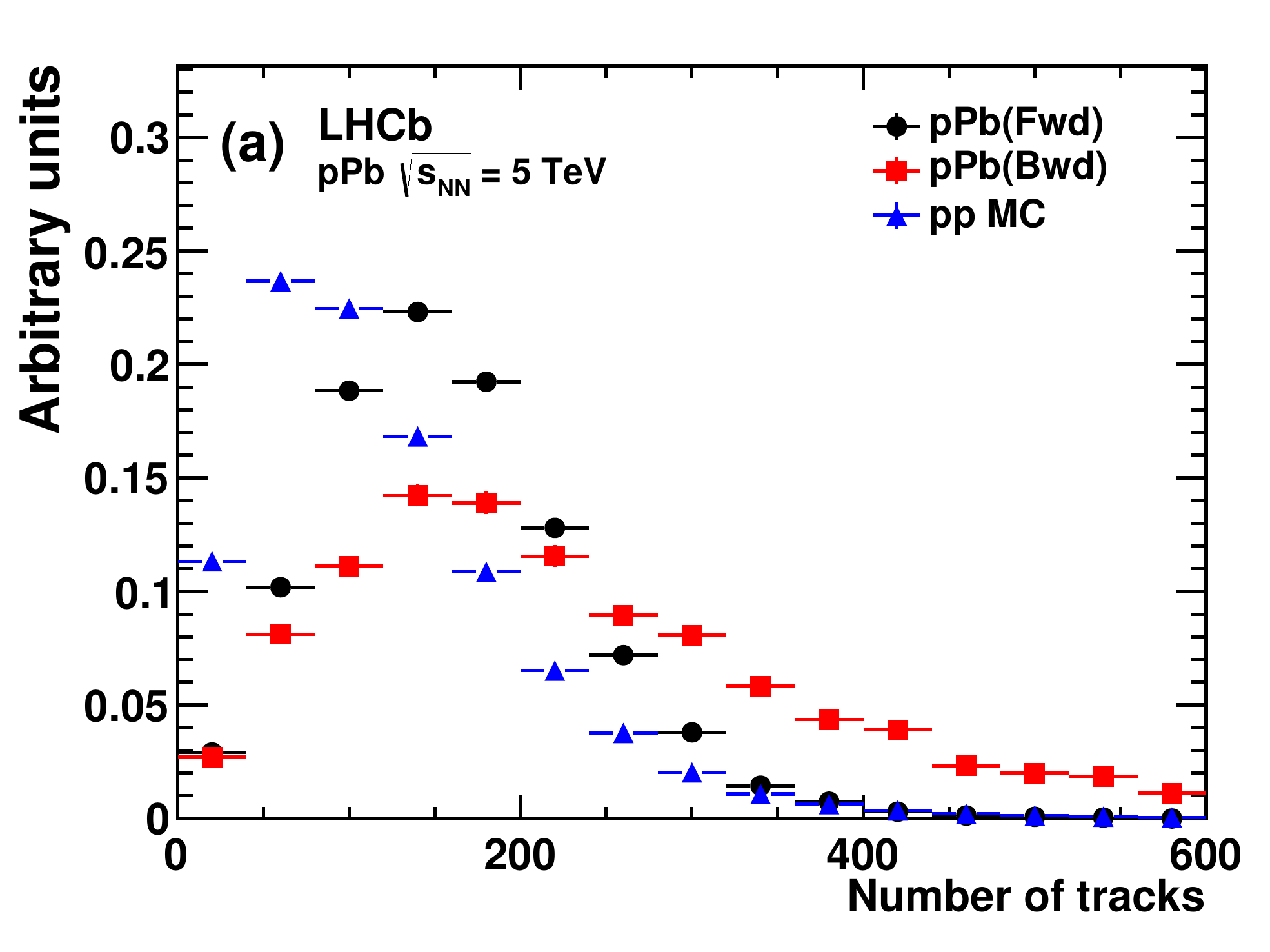}
\includegraphics[width=0.49 \textwidth]{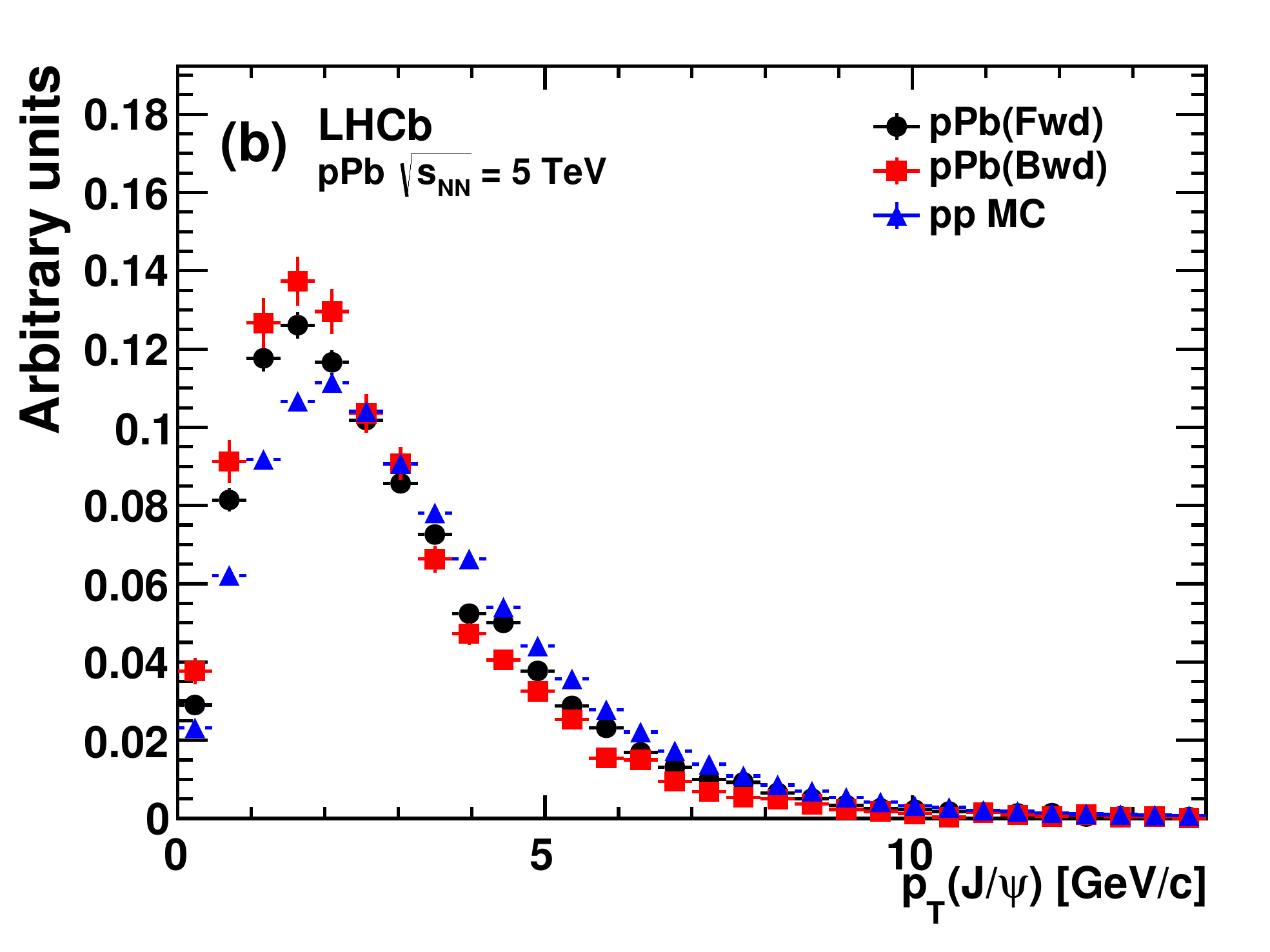}
\includegraphics[width=0.49 \textwidth]{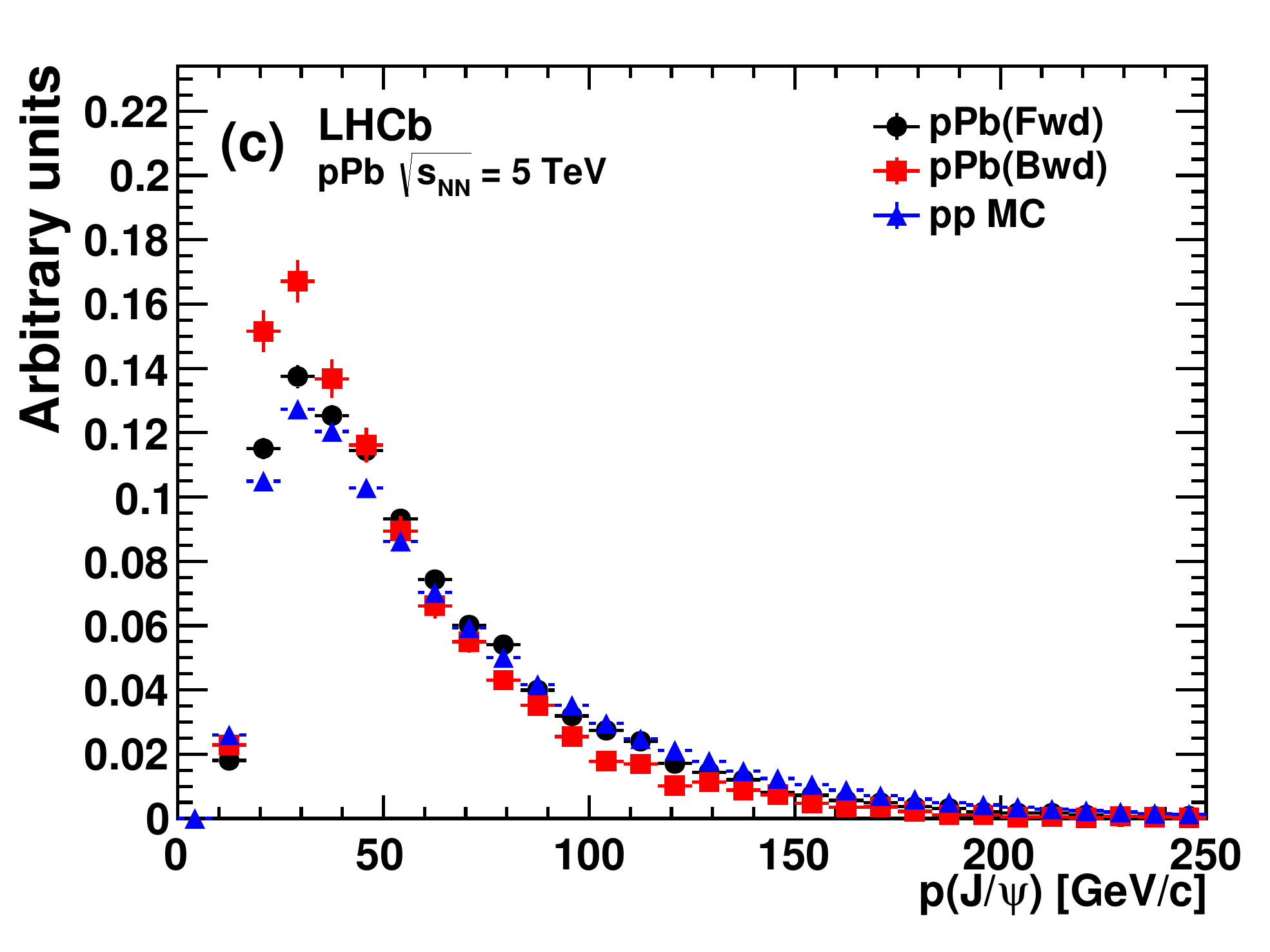}
\includegraphics[width=0.49 \textwidth]{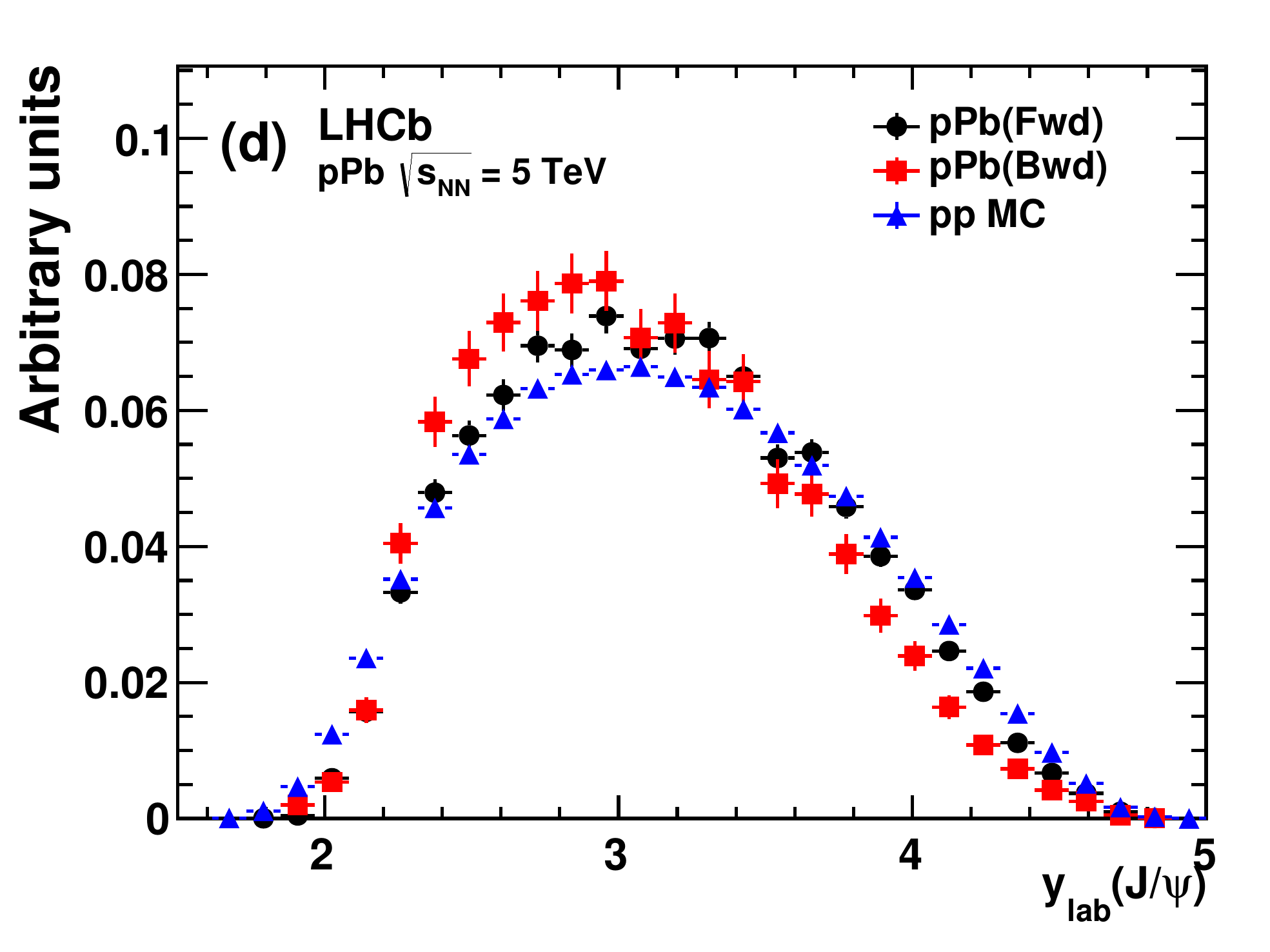}
\vspace*{-1.0cm}
\end{center}
\caption{
    \small Distributions (normalised to unitary integral) of 
        (a) track multiplicity and the $\jpsi$ 
        (b) transverse momentum $p_\mathrm{T}$,
        (c) momentum $p$, and
        (d) rapidity in laboratory frame $y_{\mathrm{lab}}$
        in (black dots) forward and (red squares) backward regions of $\pPb$ collisions,
        and in (blue triangles) simulated $pp$ collisions.
        The distributions are background subtracted using the \sPlot technique.
}
\label{fig:JpsiMCvsData}
\end{figure}

Figure~\ref{fig:JpsiMCvsData} shows the background-subtracted distributions
of the track multiplicity per event and the $\jpsi$ $\pt$,
$p$, and the rapidity in the laboratory frame $y_{\mathrm{lab}}$ 
in experimental $\pPb$ and simulated $pp$ data.
The differences in the distributions of $\pt$, $p$, and $y_\mathrm{lab}$
between data and simulated samples are small.
Sizeable differences in the distributions of the track multiplicity are observed,
particularly between the simulation and the backward sample,
for which the particle production cross-section is larger~\cite{Leitch:1999ea,Arsene:2004ux,Adler:2004eh,Adare:2010fn}.
To take this effect into account, 
the simulated $pp$ samples are reweighted to match the data 
with weight factors derived from the distributions in Fig.~\ref{fig:JpsiMCvsData}.

\section{Systematic uncertainties}
\label{sec:Systematics}
Acceptance and reconstruction efficiencies depend not only on the 
 kinematic distributions of the $\jpsi$\ meson but also on its polarisation. 
The LHCb measurement in $pp$ collisions~\cite{LHCb-PAPER-2013-008} 
indicated a longitudinal polarisation consistent with zero in most of the kinematic region.
Based on the expectation that the nuclear environment does not enhance the polarisation, 
it is assumed that the $\jpsi$\/ mesons are produced with no polarisation. 
No systematic uncertainty is assigned to the effect of polarisation in this 
analysis.

\begin{table}
  \caption{ \small 
  Relative systematic uncertainties on the differential production cross-sections.
  The uncertainty due to the radiative tail and branching fraction cancels 
  in both $R_{\pPb}$ and $R_{\mbox{\tiny{FB}}}$.
  The uncertainty due to the tracking efficiency and the luminosity partially cancels for $R_{\mbox{\tiny{FB}}}$. 
  }
\begin{center}\begin{tabular}{lcc}
    \hline
    Source                    & Forward (\%)   & Backward (\%)\\
    \hline
    \it{Correlated between bins} \\
    Mass fits                 & 2.3            &  3.4\\
    Radiative tail            & 1.0            &  1.0\\
    Muon identification       & 1.3            &  1.3\\
    Tracking efficiency       & 1.5            &  1.5\\
    Luminosity                & 1.9            &  2.1\\
    $\mathcal{B}(\jpsi\to\mu^+\mu^-)$   & 1.0 &  1.0\\
    \hline
    \it{Uncorrelated between bins} \\
    Binning                   & $0.1$ -- $8.7$   & $0.1$ -- $6.1$\\
    Multiplicity weight       & $0.1$ -- $3.0$  & $0.2$ -- $4.3$\\
    $t_z$ fit (\it{only for $\jpsi$ from $b$}) & $0.2$ -- $12\,$ & $0.2$ -- $13\,$ \\
    \hline
  \end{tabular}\end{center}
\label{tab:JpsiSystematics}
\end{table}
Several contributions to the systematic uncertainties affecting the cross-section
measurement are discussed in the following and summarised 
in Table~\ref{tab:JpsiSystematics}.
The influence of the model assumed to describe the shape of the dimuon invariant mass
distribution is estimated by adding a second Crystal Ball to the fit function.
The relative difference of $2.3\%~(3.4\%)$ in the signal yield for 
forward (backward) collisions is taken as a systematic uncertainty.
Due to the muon bremsstrahlung, 
a small fraction of signal candidates with low reconstructed invariant mass
are excluded from the signal mass region. This effect is included
in the reconstruction efficiency, and an uncertainty of $1.0\%$ is assigned
based on the comparison between the observed radiative tail in data and simulation.

The systematic uncertainties due to the muon identification efficiency 
and the track reconstruction efficiency are estimated 
using a data-driven tag-and-probe method~\cite{Jaeger:1402577}
based on partially reconstructed $\jpsi$ decays.
To estimate the uncertainty due to the muon identification efficiency,
$\jpsi$ candidates are reconstructed
with one muon identified by the muon system (``tag") and the other (``probe")
identified by selecting a track depositing the energy of a minimum-ionising particle
in the calorimeters. 
The resulting uncertainty is $1.3\%$.
Taking into account the effect of the track-multiplicity difference between 
$\pPb$ and $pp$ data, an uncertainty of $1.5\%$ is assigned
due to the track reconstruction efficiency.

From the counting rate of visible interactions in the VELO, the luminosity 
is determined with an uncertainty of $1.9\%$ ($2.1\%$) 
for the $\pPb$ forward (backward) sample. 
For both configurations the relation between visible interaction rate 
and instantaneous luminosity was calibrated 
using the van der Meer method~\cite{vanderMeer:1968zz,Burkhardt:2007zzc}. 
Details of the procedure are described in Ref.~\cite{LHCb-PAPER-2011-015}. 
The statistical uncertainties are negligible, the beam intensities are 
determined with a precision of better than $0.4\%$. 
The dominant contributions to the systematic uncertainties are $0.6\%$ ($1.3\%$) 
for the $\pPb$ forward (backward) sample due to 
the reproducibility of the van der Meer scans and uncontrolled beam drifts, 
$1.0\%$ from the absolute length scale calibration of the beam displacements, 
$0.4\%$ due to longitudinal movements of the luminous region, 
and between $0.6\%$ and $1.0\%$ from beam-beam induced background.
The uncertainty of the branching fraction of the $\jpsi\to\mup\mun$ decay
is $1.0\%$~\cite{PDG2012}.

Differences of the $p_\mathrm{T}$ and $y$ spectra between data and simulation 
within a given $(\pt,\,y)$ bin due to the finite bin sizes
can affect the result.
This effect is estimated by doubling the number of bins in $p_\mathrm{T}$ 
and shifting each rapidity bin by half a unit.
The relative difference with respect to the default binning, 
which varies between $0.1\%$ and $8.7\%$ depending on the bin,
is taken as systematic uncertainty. The uncertainties in most bins are below $2.0\%$,
but increase in the lowest rapidity bins.

To estimate the effect of reweighting the track multiplicity in the simulation,
the efficiency without reweighting is calculated.
The relative difference in each bin between the two methods 
is taken as systematic uncertainty.

Uncertainties related to the $t_z$ fit procedure are measured by
fitting directly the $t_z$ signal component, 
which is determined using the {\mbox{\em sPlot\xspace}} technique.
This gives results consistent with those obtained from the combined fit;
the relative difference between results in each bin 
is taken as systematic uncertainty.

\section{Results}
\label{sec:Results}
\begin{figure}[tb]
  \begin{center}
    \includegraphics[width=0.49 \textwidth]{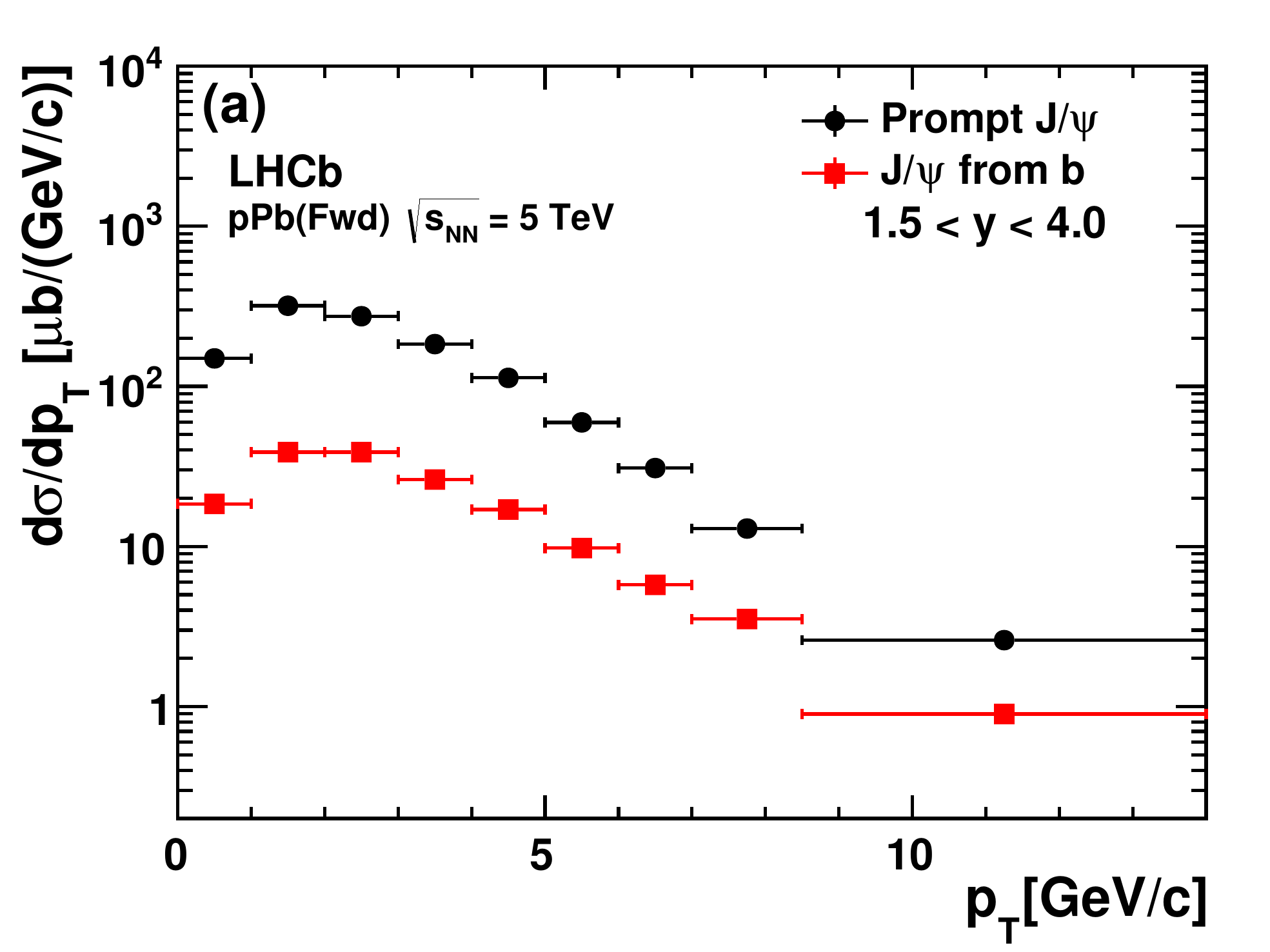}
    \includegraphics[width=0.49 \textwidth]{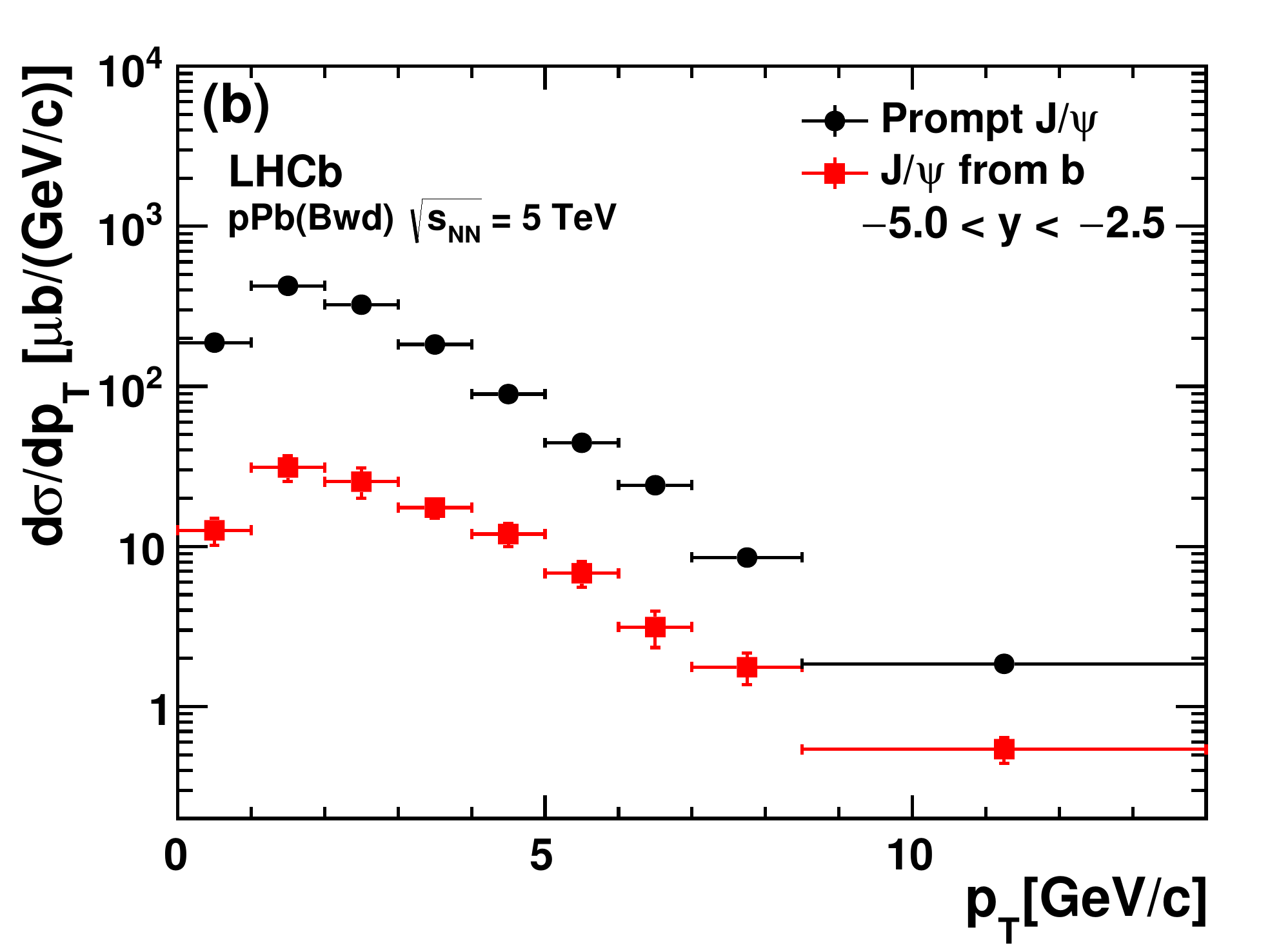}
    \includegraphics[width=0.49 \textwidth]{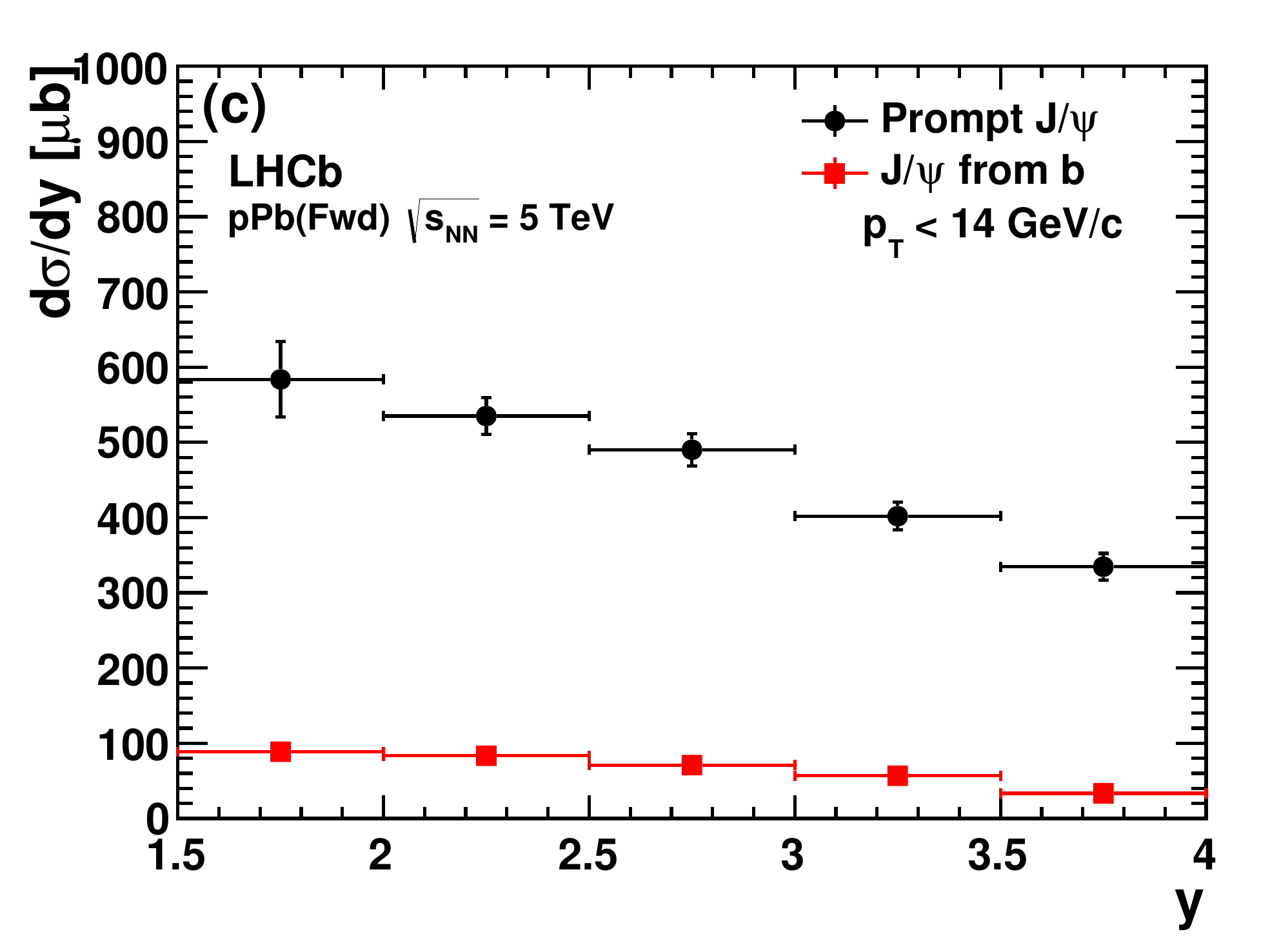}
    \includegraphics[width=0.49 \textwidth]{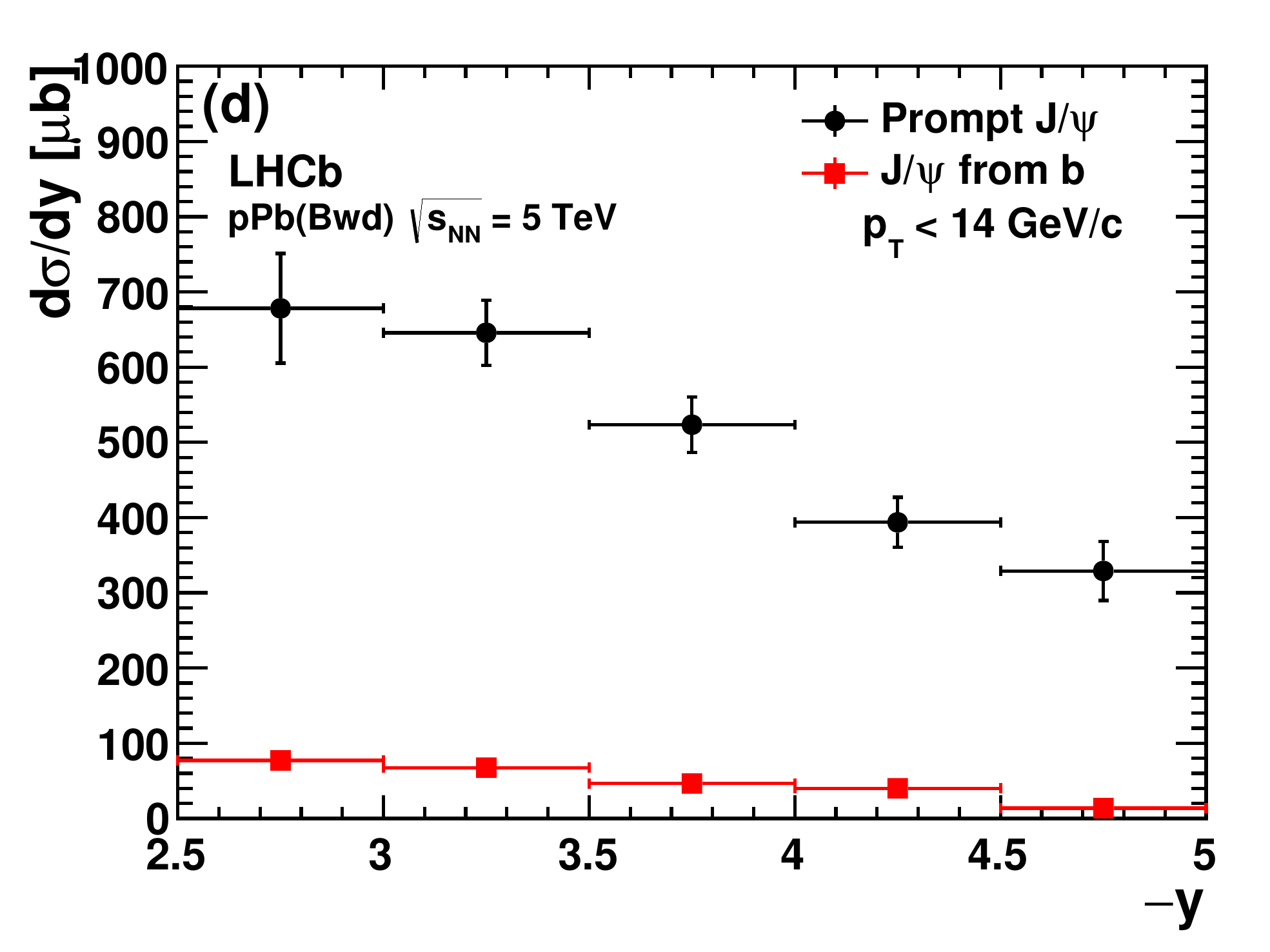}
    \vspace*{-1.0cm}
  \end{center}
  \caption{ \small 
      Single differential production cross-sections for (black dots) prompt $\jpsi$
      and (red squares) $\jpsi$ from $b$
      as functions of (a, b) $p_\mathrm{T}$ and (c, d) $y$ 
      in the (a, c) forward and (b, d) backward regions.
          }
  \label{fig:JpsiXsectionPtY_event}
\end{figure}

Single differential production cross-sections as functions of $p_\mathrm{T}$ and $y$,
for both prompt $\jpsi$ mesons and $\jpsi$ from $b$ in the $\pPb$ 
forward and backward regions, are displayed in Fig.~\ref{fig:JpsiXsectionPtY_event} and
shown in Tables~\ref{tab:JpsiPtDiffpAAp_event} and~\ref{tab:JpsiYDiffpAAp_event},
respectively, assuming no $\jpsi$ polarisation.

\begin{figure}[htb]
  \begin{center}
    \includegraphics[width=0.49 \textwidth]{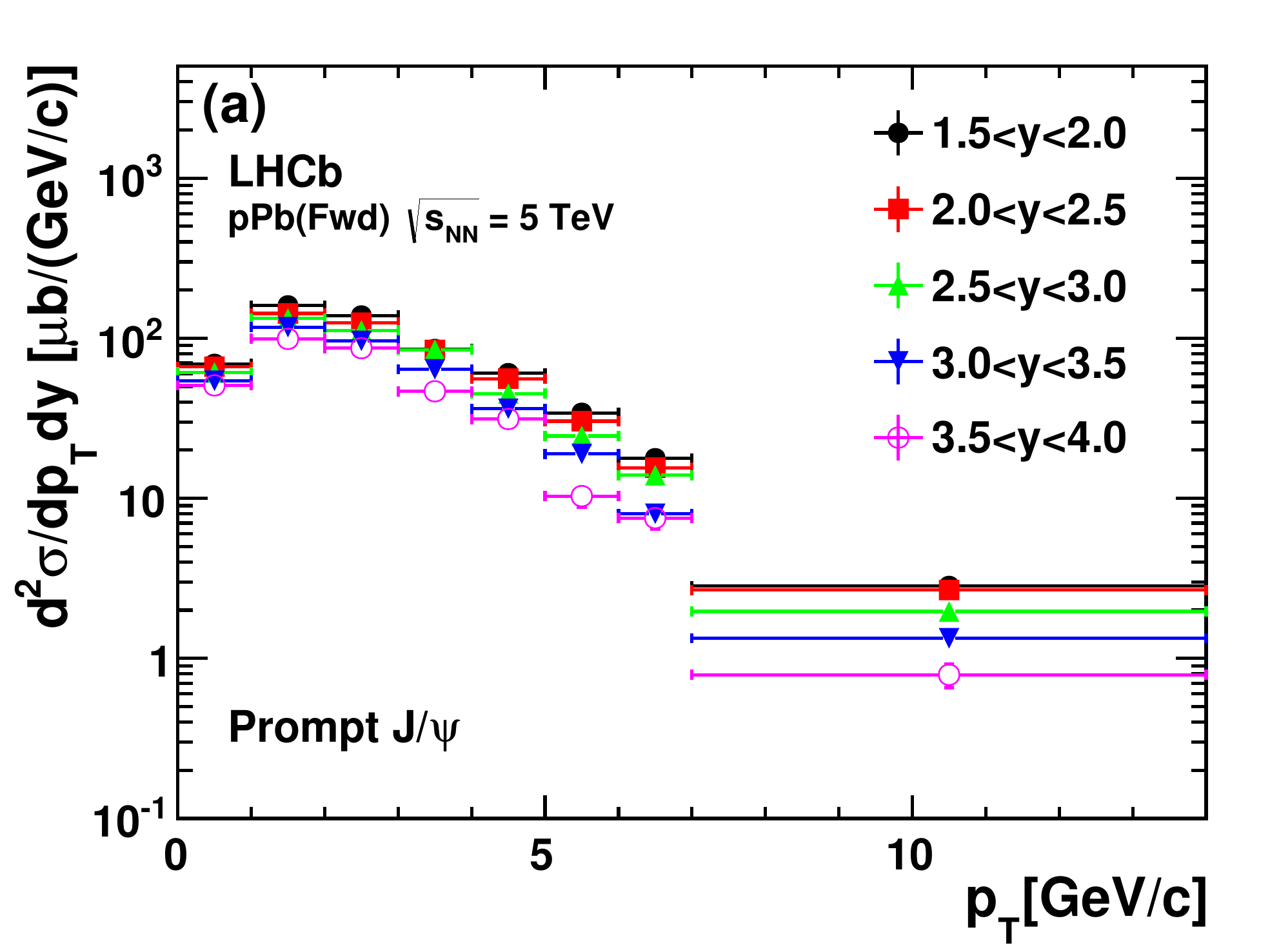}
    \includegraphics[width=0.49 \textwidth]{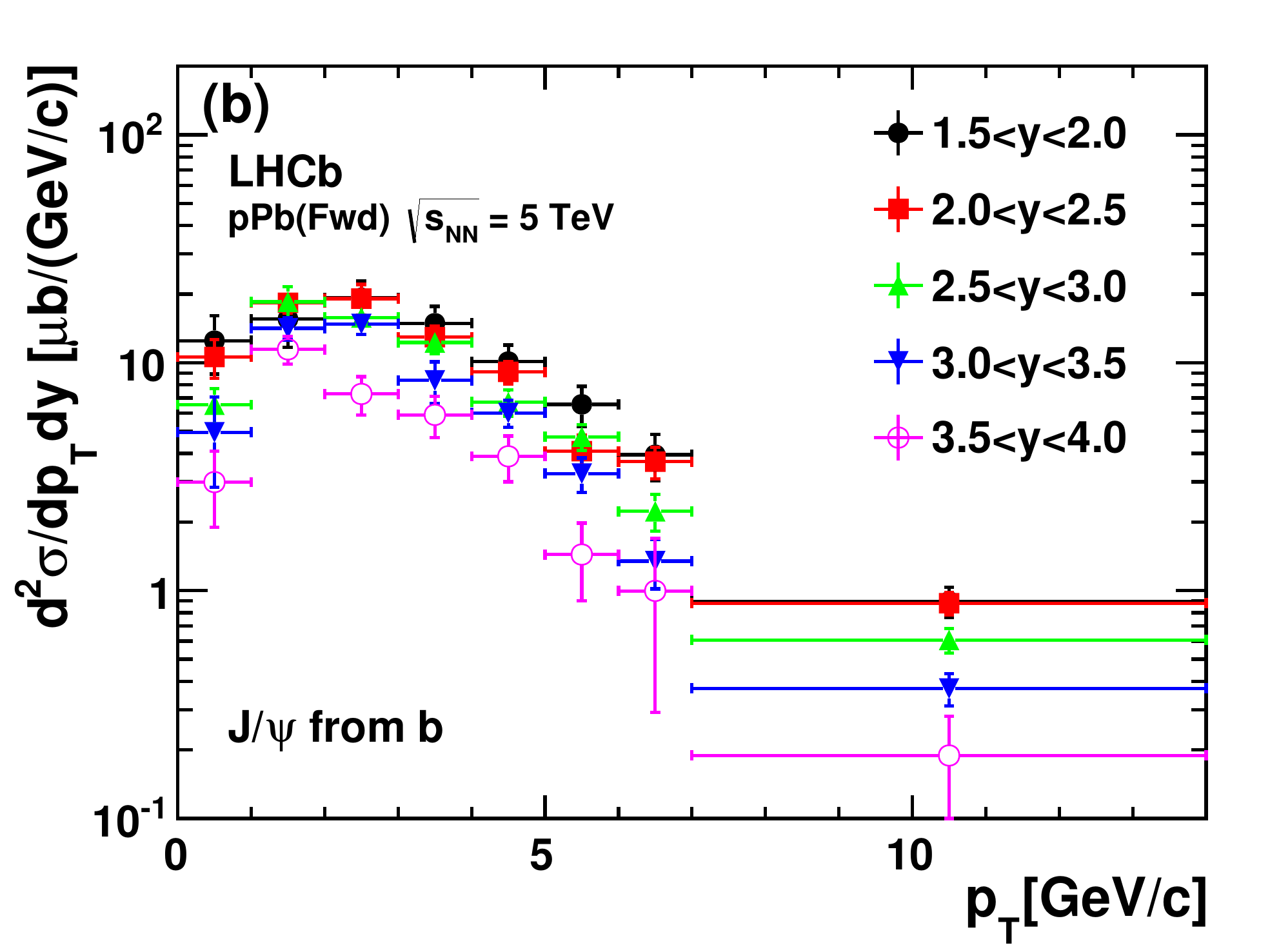}
    \vspace*{-1.0cm}
  \end{center}
  \caption{ \small 
      Double differential production cross-sections for (a) prompt $\jpsi$ mesons
      and (b) $\jpsi$ from $b$ in the forward samples.
          }
  \label{fig:JpsiDoubleDiff_event}
\end{figure}

Due to the large samples of $\pPb$ forward collisions,
the double differential production cross-sections 
can also be measured.
These results are shown in Fig.~\ref{fig:JpsiDoubleDiff_event} 
and displayed in Table~\ref{tab:JpsiDoubleDiff_event}.

The integrated production cross-sections for prompt $\jpsi$ mesons and $\jpsi$ from $b$
with $\pt<14\gevc$ in the forward and backward regions are measured to be
\begin{equation*}
\begin{split}
 &\sigma_{\mathrm{\mbox{\tiny{F}}}}(\mathrm{prompt\ }\jpsi,\,+1.5<y<+4.0)  
  =\PromptJpsipA,\\
 &\sigma_{\mathrm{\mbox{\tiny{B}}}}(\mathrm{prompt}\ \jpsi,\,-2.5<y<-5.0)
  =\PromptJpsiAp,\\
 &\sigma_{\mathrm{\mbox{\tiny{F}}}}(\jpsi\ \mathrm{from}\ b,\,\ \,+1.5<y<+4.0) 
  =\JpsiFrombpA,\\
 &\sigma_{\mathrm{\mbox{\tiny{B}}}}(\jpsi\ \mathrm{from}\ b,\,\ \,-2.5<y<-5.0)
  =\JpsiFrombAp,
\end{split}
\end{equation*}
where the first uncertainty is statistical and the second is systematic.

\begin{figure}[tb]
  \begin{center}
    \includegraphics[width=0.49 \textwidth]{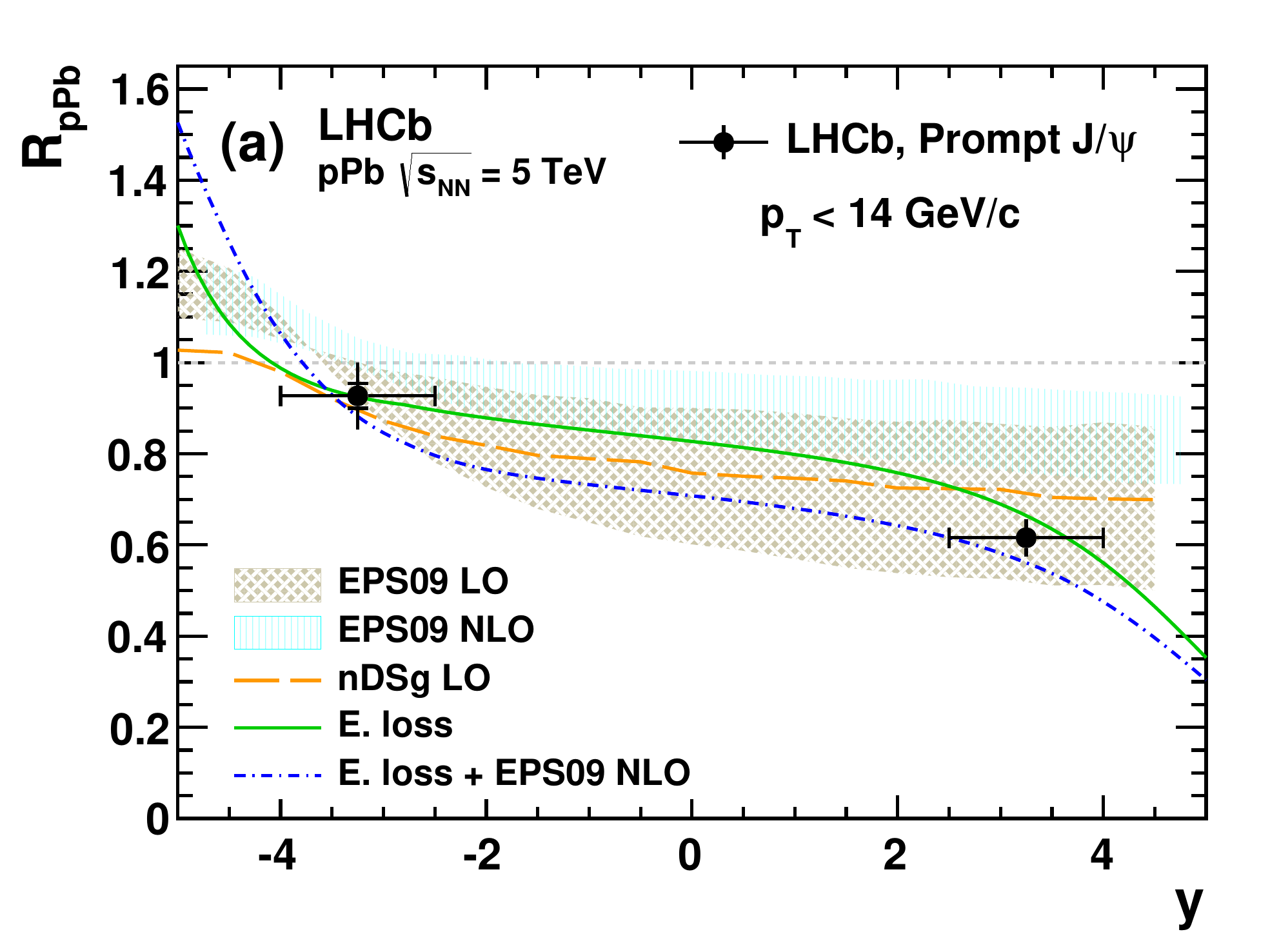}
    \includegraphics[width=0.49 \textwidth]{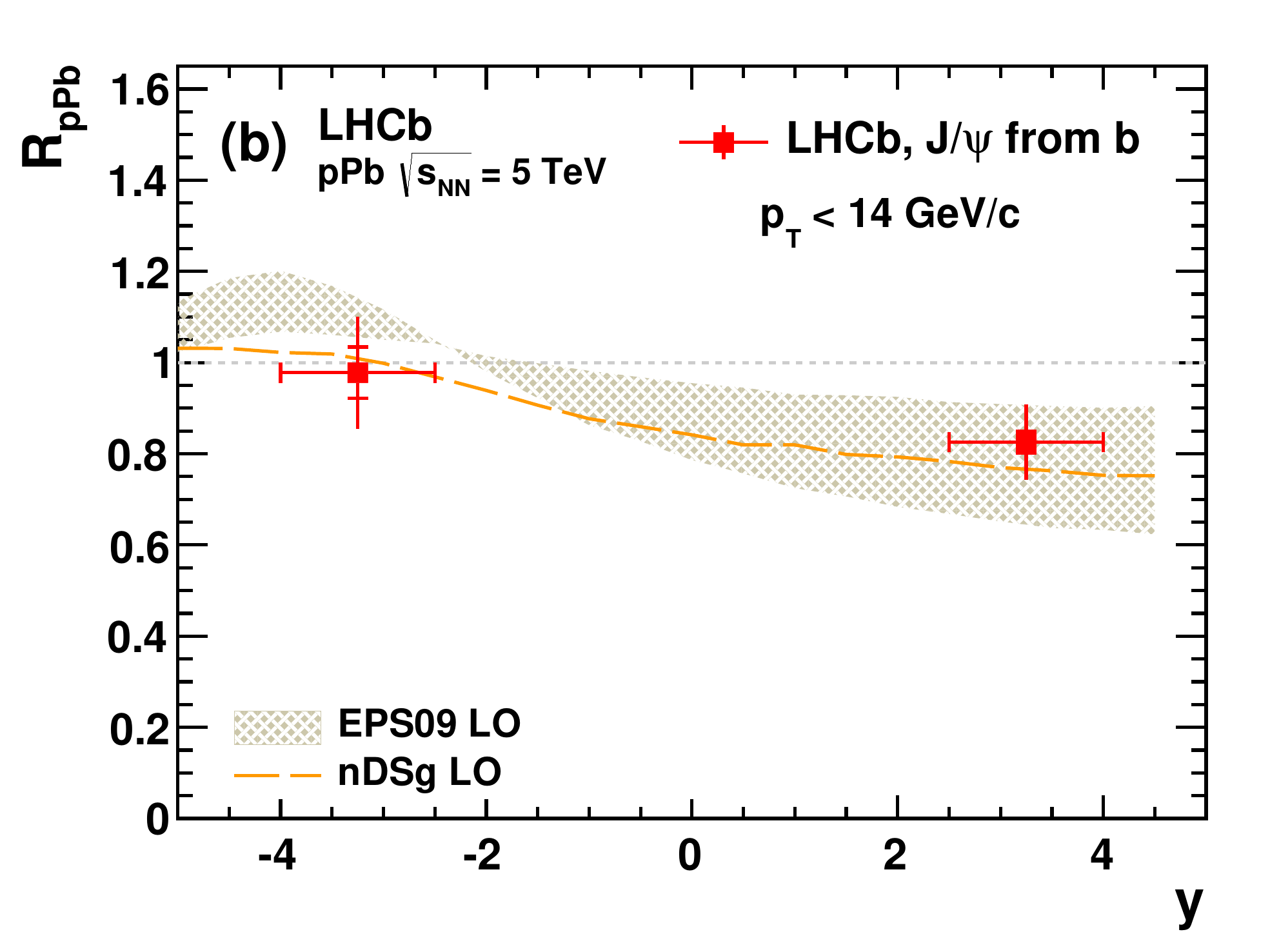}
    \vspace*{-1.0cm}
  \end{center}
  \caption{ \small 
     Nuclear modification factor $R_{\pPb}$ as a function of $y$ 
     for (a) prompt $\jpsi$ mesons and (b) $\jpsi$ from $b$,
     together with the theoretical predictions from 
     (yellow dashed line and brown band) Refs.~\cite{Ferreiro:2013pua,delValle:2014wha}, 
     (blue band) Ref.~\cite{Albacete:2013ei}, and 
     (green solid and blue dash-dotted lines) Ref.~\cite{Arleo:2012rs}.
     The inner error bars (delimited by the horizontal lines) show the statistical uncertainties; the outer ones show the statistical and systematic uncertainties added in quadrature.
     The uncertainty due to the interpolated $\jpsi$ cross-section in $pp$ collisions at $\sqrt{s}=5\tev$ is $5.5\%$ ($8.4\%$) for prompt $\jpsi$ mesosns ($\jpsi$ from $b$).
          }
  \label{fig:ModificationCompare}
\end{figure}

The $\jpsi$ production cross-section in $pp$ collisions 
at $5\mathrm{\,Te\kern -0.1em V}$, used as a reference 
to determine the nuclear modification factor $R_{\pPb}$,
is obtained by a power-law interpolation, 
$\sigma(\sqrt{s})=\left(\sqrt{s}/p_0\right)^{p_1}\,\mub$, 
of previous LHCb measurements performed
at $2.76$, $7$, and $8\tev$~\cite{LHCb-PAPER-2011-003,LHCb-PAPER-2012-039,LHCb-PAPER-2013-016}. 
For $\sqrt{s}=7$ and $8\tev$, measurements in the kinematic region
$p_\mathrm{T}<14\mathrm{\,Ge\kern -0.1em V\!/}c$ and $2.5<|y|<4.0$,
the common rapidity range of the forward and backward regions 
in the nucleon-nucleon centre-of-mass frame, are available.
The measurements at $\sqrt{s}=2.76\tev$ are rescaled to this range.
The fits give $p_0=0.67\pm0.10\tev$ and $p_1=0.49\pm0.18$ for prompt $\jpsi$ mesons,
and $p_0=1.1\pm0.2\tev$ and $p_1=10.0\pm0.8$ for $\jpsi$ from $b$.
Alternative interpolations based on linear and exponential fits are also tried;
the largest deviation from the default value is taken as a systematic uncertainty 
due to the interpolation, $3.1\%~(2.8\%)$ for prompt (from $b$) $\jpsi$ mesons.
The reference production cross-section in $pp$ collisions at $5\mathrm{\,Te\kern -0.1em V}$
for prompt $\jpsi$ mesons is $4.79\pm0.22\pm0.15~\upmu\mathrm{b}$,
and that for $\jpsi$ from $b$ is $0.47\pm0.04\pm0.01~\upmu\mathrm{b}$~\cite{CommonNote}.
The nuclear modification factor $R_{\pPb}$ is then determined 
in the rapidity ranges $-4.0<y<-2.5$ and $2.5<y<4.0$ for both prompt $\jpsi$ mesons
and $\jpsi$ from $b$.
Figure~\ref{fig:ModificationCompare}(a) shows the nuclear modification factor
for prompt $\jpsi$ production, 
together with several theoretical predictions~\cite{Arleo:2012rs,Ferreiro:2013pua,Albacete:2013ei}.
Calculations in Ref.~\cite{Ferreiro:2013pua} are based on 
the Leading Order Colour Singlet Model (LO CSM)~\cite{Chang:1979nn,Baier:1981uk}, 
taking into account the modification effects of the gluon distribution function
in nuclei with the parameterisation EPS09~\cite{Eskola:2009uj} or nDSg~\cite{deFlorian:2003qf}.
The Next-to-Leading Order Colour Evaporation Model
(NLO CEM)~\cite{CEM} is used in Ref.~\cite{Albacete:2013ei}, 
considering the parton shadowing with EPS09 parameterisation.
Reference~\cite{Arleo:2012rs} provides theoretical predictions of a coherent 
parton energy loss effect both in initial and final states, with or without additional parton shadowing effects parameterised with EPS09.
The single free parameter $q_0$ in this model is $0.055~(0.075)~{\gev}^2/\fm$ when EPS09 is (not) taken into account.
A suppression of about $40\%$ at large rapidity 
is observed for prompt $\jpsi$ production.
The measurements agree with most predictions.
However, the calculation~\cite{Albacete:2013ei} based on 
NLO CEM with the EPS09 parameterisation
provides a less good description of the measurement in the forward region.
Figure~\ref{fig:ModificationCompare}(b) shows the nuclear modification factor
for $\jpsi$ from $b$, together with the theoretical predictions~\cite{delValle:2014wha}.
The data show a modest suppression of $\jpsi$ from $b$ production 
in $\pPb$ forward region,
with respect to that in $pp$ collisions.
This is the first indication of the suppression of $b$ hadron production in $\pPb$ collisions.
The theoretical predictions agree with the measurement in the forward region. 
In the backward region the agreement is not as good.
The observed production suppression of $\jpsi$ from $b$ with respect to $pp$ collisions
is smaller than that of prompt $\jpsi$, which is consistent with theoretical predictions.
The measured values of the nuclear modification factor,
together with the results for inclusive $\jpsi$ mesons, 
are given in Table~\ref{tab:RpA}.

\begin{figure}[tb]
  \begin{center}
    \includegraphics[width=0.49 \textwidth]{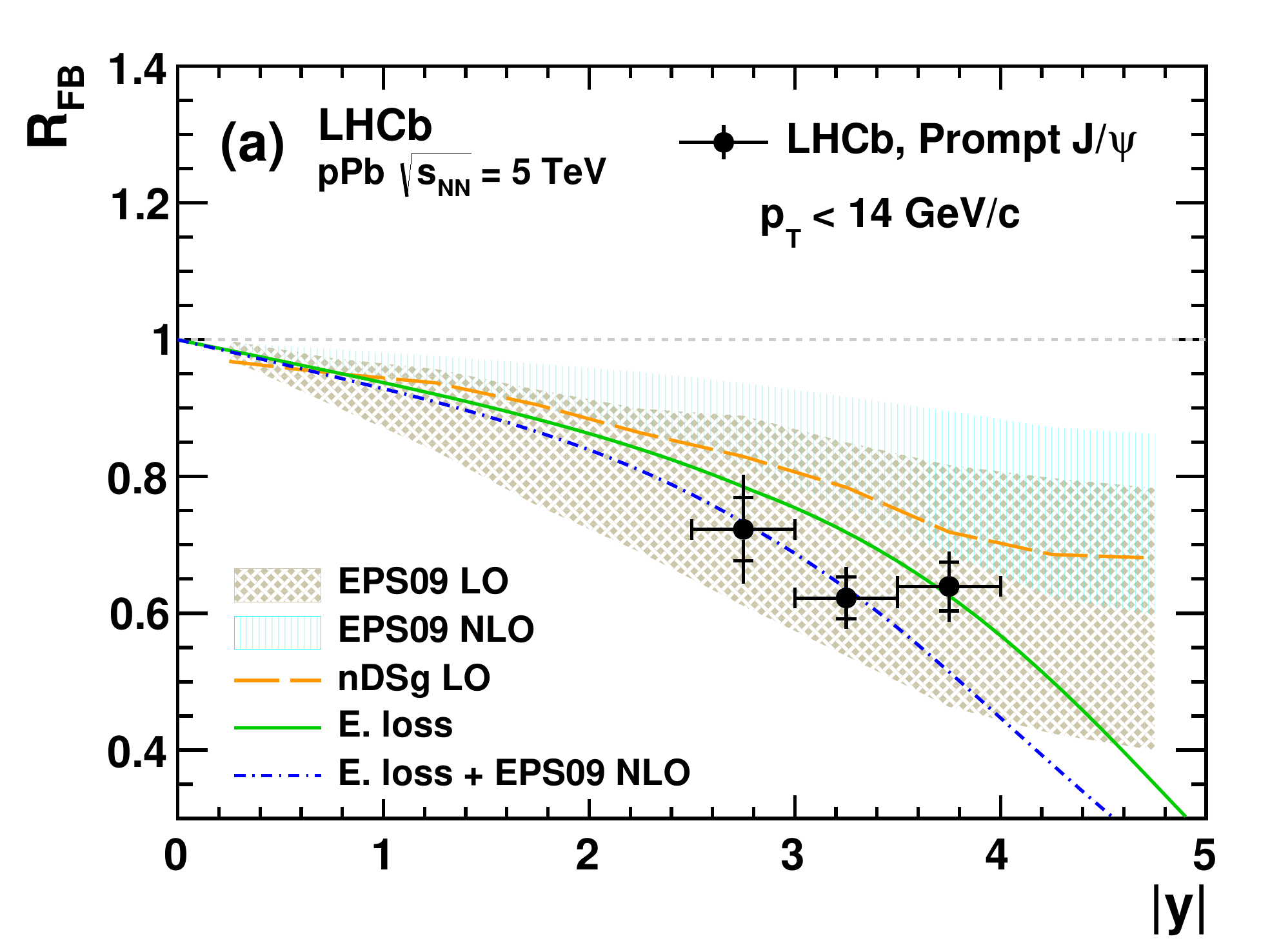}
    \includegraphics[width=0.49 \textwidth]{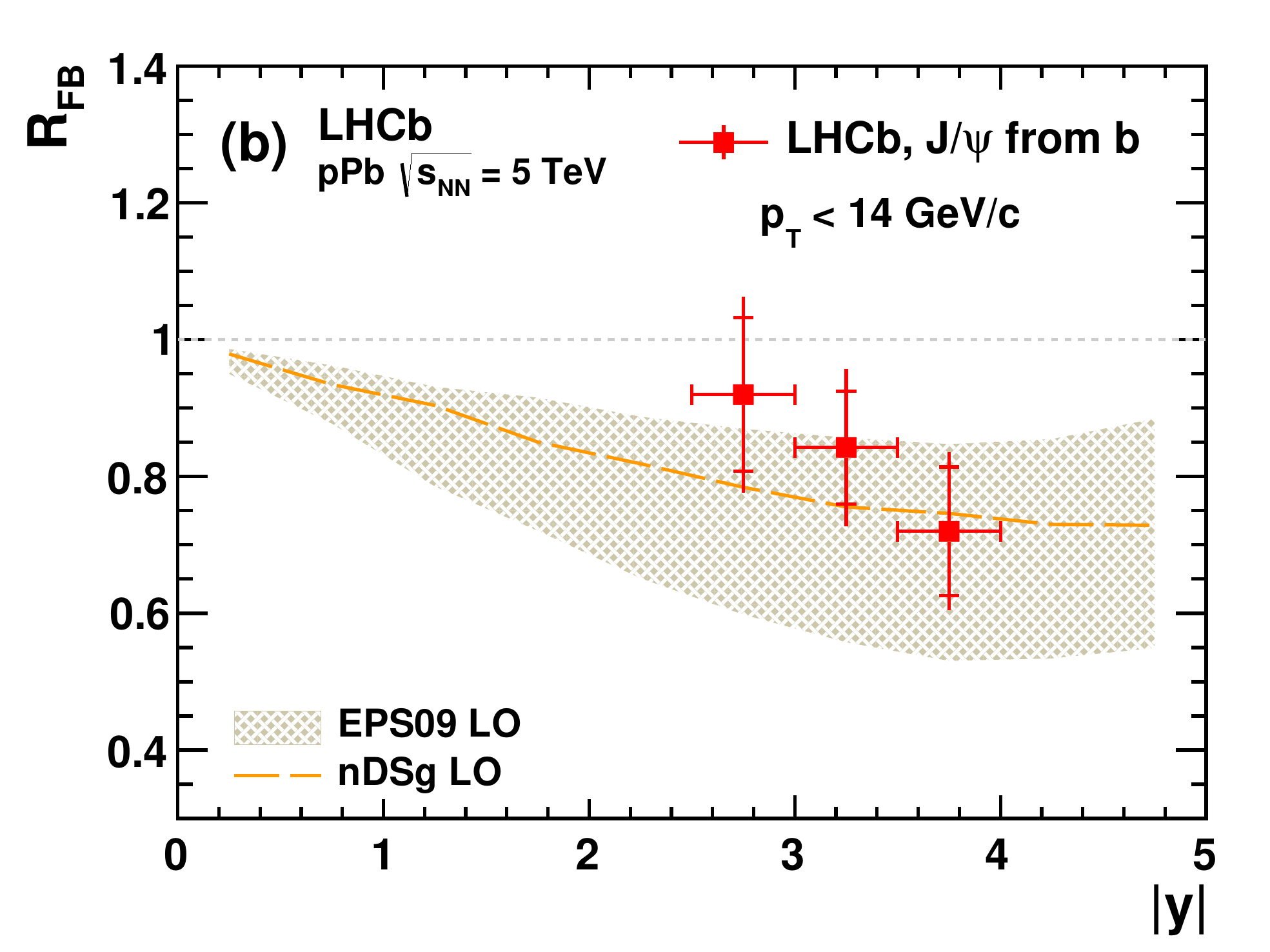}
    \vspace*{-1.0cm}
  \end{center}
  \caption{ \small 
     Forward-backward production ratio $R_{\mbox{\tiny{FB}}}$ as a function of $|y|$
     for (a) prompt $\jpsi$ mesons and (b) $\jpsi$ from $b$,
     together with the theoretical predictions from
     (yellow dashed line and brown band) Refs.~\cite{Ferreiro:2013pua,delValle:2014wha}, 
     (blue band) Ref.~\cite{Albacete:2013ei}, and 
     (green solid and blue dash-dotted lines) Ref.~\cite{Arleo:2012rs}.
     The inner error bars (delimited by the horizontal lines) show the statistical uncertainties; the outer ones show the statistical and systematic uncertainties added in quadrature.
          }
  \label{fig:R_FB}
\end{figure}

\begin{figure}[tb]
  \begin{center}
    \includegraphics[width=0.49 \textwidth]{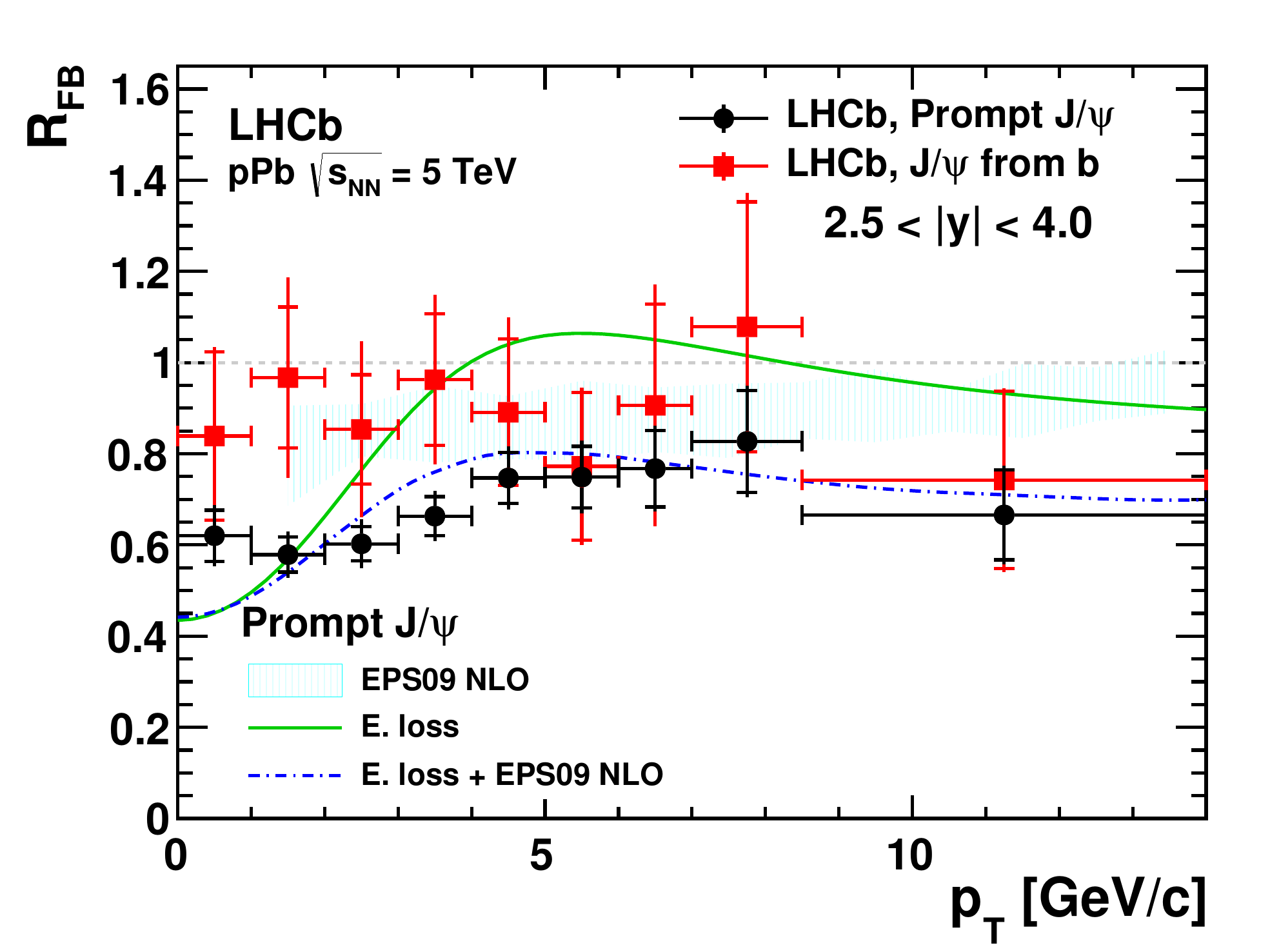}
    \vspace*{-1.0cm}
  \end{center}
  \caption{ \small 
     Forward-backward production ratio $R_{\mbox{\tiny{FB}}}$
     for (black dot) prompt $\jpsi$ mesons and (red square) $\jpsi$ from $b$
     as a function of $p_\mathrm{T}$ in the rapidity range $2.5<|y|<4.0$.
     The theoretical predictions from
     (blue band) Ref.~\cite{Albacete:2013ei} 
     and (green solid and blue dash-dotted lines) Ref.~\cite{Arleo:2013zua}
     are for prompt $\jpsi$ mesons.
     The inner error bars (delimited by the horizontal lines) show the statistical uncertainties; the outer ones show the statistical and systematic uncertainties added in quadrature.
          }
  \label{fig:R_FB_pt}
\end{figure}

Figure~\ref{fig:R_FB} shows the forward-backward production ratio $R_{\mbox{\tiny{FB}}}$ 
as a function of $|y|$, compared with 
theoretical calculations~\cite{Arleo:2012rs,Ferreiro:2013pua,Albacete:2013ei,delValle:2014wha}.
The value of $R_{\mbox{\tiny{FB}}}$ for $\jpsi$ from $b$ is closer to unity 
than for prompt $\jpsi$ mesons,
indicating a smaller asymmetry in the forward-backward production.
The results agree with theoretical predictions.
The calculation~\cite{Albacete:2013ei} with the EPS09 NLO nPDF alone predicts 
a smaller forward-backward production asymmetry for prompt $\jpsi$ mesons than observed.
Figure~\ref{fig:R_FB_pt} shows the forward-backward production ratio 
$R_{\mbox{\tiny{FB}}}$ as a function of $p_\mathrm{T}$
for prompt $\jpsi$ mesons and $\jpsi$ from $b$
in the range $2.5<y<4.0$ of the nucleon-nucleon centre-of-mass frame.
Theoretical predictions~\cite{Albacete:2013ei,Arleo:2013zua} are only available
for prompt $\jpsi$ mesons. 
The calculation~\cite{Arleo:2013zua} based on 
parton energy loss with the EPS09 NLO nPDF agrees with the measurement 
of $R_{\mbox{\tiny{FB}}}$ for prompt $\jpsi$ mesons. 
The measured values of the forward-backward production ratio $R_{\mbox{\tiny{FB}}}$
are given in Tables~\ref{tab:RFB_y} and \ref{tab:RFB_pt},
where the results for inclusive $\jpsi$ mesons are also listed.

\section{Conclusion}
\label{sec:Conclusion}
  The production of prompt $\jpsi$ mesons and of $\jpsi$ from $b$-hadron decays
  is studied in $\pPb$ collisions
  with the {\mbox{LHCb}\xspace}\ detector at the nucleon-nucleon centre-of-mass energy 
  $\sqrt{s_{\mbox{\tiny{\it NN}}}}=5\mathrm{\,Te\kern -0.1em V}$. 
  The measurement is performed as a function of the transverse momentum
  and rapidity of the $\jpsi$ meson
  in the region $p_\mathrm{T}<14\mathrm{\,Ge\kern -0.1em V\!/}c$\/ 
  and $1.5<y<4.0$\/ (forward) and $-5.0<y<-2.5$\/ (backward). 
  The nuclear modification factor $R_{\pPb}$ 
  and the forward-backward production ratio $R_{\mbox{\tiny{FB}}}$\/ 
  are determined for the first time separately for prompt $\jpsi$ mesons 
  and those from $b$-hadron decays.
  The measurement indicates that cold nuclear matter effects are less pronounced 
  for $\jpsi$ mesons from $b$-hadron decays, hence for $b$ hadrons, 
  than for prompt $\jpsi$ mesons.
  These results show good agreement with the available theoretical predictions
  and provide useful constraints to the parameterisation of theoretical models.
  The measured nuclear modification factor for prompt $\jpsi$ mesons shows that
  it is necessary to include cold nuclear matter effects in the interpretation
  of quark-gluon plasma signatures in heavy-ion collisions.
  The results for inclusive $\jpsi$ mesons are in agreement with those
  presented by the ALICE collaboration~\cite{ALICE-Jpsi-in-pA}.


\section*{Acknowledgements}

\noindent 
We are grateful for useful discussions with the ALICE collaboration.
We wish to thank also F. Arleo, J. P. Lansberg, and R. Vogt 
for stimulating and helpful suggestions. 
We express our gratitude to our colleagues in the CERN
accelerator departments for the excellent performance of the LHC. We
thank the technical and administrative staff at the LHCb
institutes. We acknowledge support from CERN and from the national
agencies: CAPES, CNPq, FAPERJ and FINEP (Brazil); NSFC (China);
CNRS/IN2P3 and Region Auvergne (France); BMBF, DFG, HGF and MPG
(Germany); SFI (Ireland); INFN (Italy); FOM and NWO (The Netherlands);
SCSR (Poland); MEN/IFA (Romania); MinES, Rosatom, RFBR and NRC
``Kurchatov Institute'' (Russia); MinECo, XuntaGal and GENCAT (Spain);
SNSF and SER (Switzerland); NAS Ukraine (Ukraine); STFC (United
Kingdom); NSF (USA). We also acknowledge the support received from the
ERC under FP7. The Tier1 computing centres are supported by IN2P3
(France), KIT and BMBF (Germany), INFN (Italy), NWO and SURF (The
Netherlands), PIC (Spain), GridPP (United Kingdom). We are thankful
for the computing resources put at our disposal by Yandex LLC
(Russia), as well as to the communities behind the multiple open
source software packages that we depend on.

\clearpage
{\noindent\bf\Large Appendices}

\appendix

\section{Results in tables}
\label{App:Results}
\begin{table}[!htb]
  \caption{ \small 
    Single differential production cross-sections (in $\mub/(\gevc)$) 
    for prompt \jpsi\ mesons and \jpsi\ from $b$ as functions of transverse momentum.
    The first uncertainty is statistical, 
    the second is the component of the systematic uncertainty
    that is uncorrelated between bins, and the third is the correlated component.
  }
 \begin{center}
 \scalebox{1.0}{%
   \begin{tabular}{lcc}
   \toprule
\pt [\gevc] & $\deriv\sigma/\deriv\pt$ (prompt \jpsi) & $\deriv\sigma/\deriv\pt$ (\jpsi from $b$)\\
 \midrule
 \multicolumn{3}{l}{Forward $(1.5<y<4.0)$ }\\
 \midrule
 $ 0.0-1.0$  & $149.6    \pm\xx5.7\pm2.0\pm\xx5.8$ & $  18.5\pm1.6\pm0.9\pm0.7$ \\
 $ 1.0-2.0$  & $319.0    \pm  11.1\pm5.3\pm  12.4$ & $  38.9\pm2.3\pm0.5\pm1.5$ \\
 $ 2.0-3.0$  & $274.4    \pm\xx7.1\pm4.6\pm  10.7$ & $  38.8\pm2.1\pm0.9\pm1.5$ \\
 $ 3.0-4.0$  & $183.7    \pm\xx5.1\pm3.4\pm\xx7.1$ & $  26.2\pm1.6\pm1.8\pm1.0$ \\
 $ 4.0-5.0$  & $113.0    \pm\xx3.2\pm1.7\pm\xx4.4$ & $  17.0\pm1.1\pm0.2\pm0.7$ \\
 $ 5.0-6.0$  & $\xx59.6  \pm\xx2.1\pm0.7\pm\xx2.3$ & $\xx9.8\pm0.8\pm0.3\pm0.4$\\
 $ 6.0-7.0$  & $\xx30.9  \pm\xx1.4\pm0.2\pm\xx1.2$ & $\xx5.8\pm0.6\pm0.2\pm0.2$\\
 $ 7.0-8.5$  & $\xx12.9  \pm\xx0.6\pm0.2\pm\xx0.5$ & $\xx3.5\pm0.3\pm0.0\pm0.1$\\
 $ 8.5-14$ & $\xx\xx2.6\pm\xx0.1\pm0.0\pm\xx0.1$ & $\xx0.9\pm0.1\pm0.0\pm0.0$\\
\midrule
 \multicolumn{3}{l}{Backward $(-5.0<y<-2.5)$ }\\
\midrule
 $ 0.0 - 1.0$  & $    187.7\pm  14.0\pm5.6\pm\xx9.0$ & $  12.6\pm2.1\pm1.0\pm0.6$ \\
 $ 1.0 - 2.0$  & $    425.0\pm  23.6\pm7.8\pm  20.5$ & $  31.1\pm3.8\pm4.1\pm1.5$ \\
 $ 2.0 - 3.0$  & $    323.9\pm  16.7\pm9.1\pm  15.6$ & $  25.4\pm3.0\pm4.4\pm1.2$ \\
 $ 3.0 - 4.0$  & $    182.7\pm\xx9.6\pm2.2\pm\xx8.8$ & $  17.5\pm2.1\pm1.1\pm0.8$ \\
 $ 4.0 - 5.0$  & $\xx  89.6\pm\xx5.5\pm1.3\pm\xx4.3$ & $  12.0\pm1.7\pm0.8\pm0.6$ \\
 $ 5.0 - 6.0$  & $\xx  44.4\pm\xx3.1\pm1.0\pm\xx2.1$ & $\xx6.8\pm1.2\pm0.2\pm0.3$ \\
 $ 6.0 - 7.0$  & $\xx  24.1\pm\xx2.1\pm0.4\pm\xx1.2$ & $\xx3.1\pm0.7\pm0.4\pm0.2$ \\
 $ 7.0 - 8.5$  & $\xx\xx8.5\pm\xx0.9\pm0.3\pm\xx0.4$ & $\xx1.8\pm0.4\pm0.1\pm0.1$ \\
 $ 8.5 - 14 $  & $\xx\xx1.9\pm\xx0.2\pm0.0\pm\xx0.1$ & $\xx0.5\pm0.1\pm0.0\pm0.0$ \\
 \bottomrule
 \end{tabular}%
 }\end{center}
\label{tab:JpsiPtDiffpAAp_event}
\end{table}
\begin{table}[tb]
  \caption{ \small 
    Single differential production cross-sections (in \mub) for prompt \jpsi\ mesons
    and \jpsi\ from $b$ as functions of rapidity.
    The first uncertainty is statistical, the second is the component of the systematic uncertainty
    that is uncorrelated between bins, and the third is the correlated component.
  }
 \begin{center}
 \scalebox{1.0}{%
   \begin{tabular}{lcc}
       \toprule
$|y|$ & $\deriv\sigma/\deriv{}y$ (prompt \jpsi) & $\deriv\sigma/\deriv{}y$ (\jpsi from $b$) \\
 \midrule
 \multicolumn{3}{l}{Forward $(\pt<14\gevc)$ }\\
 \midrule
 $1.5-2.0$ & $583.7\pm  21.0\pm  39.6\pm22.7$ & $88.8\pm5.9\pm6.1\pm3.4$\\
 $2.0-2.5$ & $535.0\pm  11.8\pm\xx5.6\pm20.8$ & $83.4\pm3.4\pm2.1\pm3.2$\\
 $2.5-3.0$ & $490.2\pm\xx9.7\pm\xx2.6\pm19.0$ & $71.0\pm2.8\pm0.4\pm2.8$\\
 $3.0-3.5$ & $401.9\pm\xx8.5\pm\xx5.2\pm15.6$ & $56.9\pm2.6\pm2.4\pm2.2$\\
 $3.5-4.0$ & $334.7\pm\xx8.6\pm\xx8.4\pm13.0$ & $33.4\pm2.6\pm2.5\pm1.3$\\
 \midrule
 \multicolumn{3}{l}{Backward $(\pt<14\gevc)$ }\\
 \midrule
 $2.5-3.0$ & $678.2\pm41.8\pm  50.6\pm32.6$ & $77.2\pm8.9\pm6.4\pm3.7$\\
 $3.0-3.5$ & $645.7\pm28.7\pm\xx9.6\pm31.1$ & $67.5\pm5.8\pm4.7\pm3.3$\\
 $3.5-4.0$ & $523.4\pm25.9\pm\xx8.0\pm25.2$ & $46.5\pm4.8\pm1.1\pm2.2$\\
 $4.0-4.5$ & $393.7\pm26.1\pm\xx8.7\pm18.9$ & $40.1\pm5.2\pm4.6\pm1.9$\\
 $4.5-5.0$ & $329.0\pm31.3\pm  15.7\pm15.8$ & $13.8\pm4.5\pm1.3\pm0.7$\\
 \bottomrule
  \end{tabular}%
 }\end{center}
\label{tab:JpsiYDiffpAAp_event}
\end{table}

\begin{sidewaystable}[tb]  
\caption{ \small 
    Double differential production cross-sections (in ${\rm \mub}/(\gevc)$) 
    for prompt \jpsi\ mesons and \jpsi\ from $b$ 
    as functions of \pt and $y$ in $\pPb$ forward data. 
    The first uncertainty is statistical, the second is the component of  
    the systematic uncertainty that is uncorrelated between bins, 
    and the third is the correlated component.
  }
 \begin{small}
 \scalebox{0.9}{%
   \begin{tabular}{@{}lccccc@{}}
   \toprule
\pt [\gevc] & $1.5<y<2.0$ & $2.0<y<2.5$ & $2.5<y<3.0$ & $3.0<y<3.5$ & $3.5<y<4.0$ \\
 \midrule
 prompt \jpsi\ & &&&&\\
 \midrule
 $0.0-1.0$ & $ \xx69.1\pm\xx7.7\pm\xx3.7 \pm2.7$ & $  \xx66.3\pm4.5\pm0.7\pm2.6$ & $  \xx61.1\pm4.0\pm0.8\pm2.4$ & $  \xx54.2\pm3.3\pm1.9\pm2.1$ & $  50.7\pm3.3\pm2.3\pm2.0$ \\
 $1.0-2.0$ & $   160.2\pm  11.9\pm  12.4 \pm6.2$ & $    142.7\pm6.9\pm1.6\pm5.5$ & $    132.8\pm5.5\pm1.5\pm5.2$ & $    117.3\pm4.9\pm0.9\pm4.6$ & $  99.1\pm4.9\pm2.4\pm3.8$ \\
 $2.0-3.0$ & $   138.5\pm\xx9.9\pm  10.0 \pm5.4$ & $    125.2\pm5.4\pm1.4\pm4.9$ & $    111.8\pm4.4\pm0.6\pm4.3$ & $  \xx96.3\pm4.0\pm1.4\pm3.7$ & $  86.7\pm4.9\pm4.1\pm3.4$ \\
 $3.0-4.0$ & $ \xx85.6\pm\xx7.0\pm\xx7.6 \pm3.3$ & $  \xx84.5\pm3.9\pm1.2\pm3.3$ & $  \xx84.7\pm3.4\pm1.9\pm3.3$ & $  \xx64.0\pm2.9\pm1.6\pm2.5$ & $  46.6\pm3.5\pm0.7\pm1.8$ \\
 $4.0-5.0$ & $ \xx60.4\pm\xx4.5\pm\xx4.0 \pm2.3$ & $  \xx55.9\pm2.6\pm0.9\pm2.2$ & $  \xx45.1\pm2.0\pm0.9\pm1.8$ & $  \xx36.3\pm1.9\pm0.4\pm1.4$ & $  31.3\pm2.4\pm0.6\pm1.2$ \\
 $5.0-6.0$ & $ \xx34.2\pm\xx3.0\pm\xx0.9 \pm1.3$ & $  \xx30.3\pm1.7\pm0.8\pm1.2$ & $  \xx24.5\pm1.3\pm0.5\pm1.0$ & $  \xx18.9\pm1.2\pm0.3\pm0.7$ & $  10.3\pm1.4\pm0.5\pm0.4$ \\
 $6.0-7.0$ & $ \xx17.8\pm\xx1.9\pm\xx0.6 \pm0.7$ & $  \xx15.5\pm1.1\pm0.1\pm0.6$ & $  \xx14.0\pm0.9\pm0.1\pm0.5$ & $\xx\xx8.0\pm0.8\pm0.1\pm0.3$ & $\xx7.5\pm1.0\pm0.2\pm0.3$ \\
 $7.0-14$ & $\xx\xx2.8\pm\xx0.2\pm\xx0.1 \pm0.1$ & $\xx\xx2.7\pm0.2\pm0.1\pm0.1$ & $\xx\xx2.0\pm0.1\pm0.0\pm0.1$ & $\xx\xx1.3\pm0.1\pm0.0\pm0.1$ & $\xx0.8\pm0.1\pm0.0\pm0.0$ \\
 \midrule
 \jpsi\ from $b$ & &&&&\\
 \midrule
 $0.0-1.0$ & $  12.5\pm2.7\pm2.4\pm0.5$ & $  10.6\pm1.3\pm1.5\pm0.4$ & $\xx6.5\pm1.0\pm0.6\pm0.3$ & $\xx5.0\pm0.9\pm1.9\pm0.2$ & $\xx3.0\pm0.8\pm0.8\pm0.1$ \\
 $1.0-2.0$ & $  15.6\pm2.7\pm2.7\pm0.6$ & $  18.3\pm1.7\pm0.3\pm0.7$ & $  18.5\pm1.5\pm2.5\pm0.7$ & $  14.2\pm1.4\pm0.4\pm0.5$ & $  11.4\pm1.5\pm0.5\pm0.4$ \\
 $2.0-3.0$ & $  19.3\pm2.9\pm1.9\pm0.7$ & $  19.1\pm1.6\pm2.3\pm0.7$ & $  15.8\pm1.3\pm0.5\pm0.6$ & $  14.8\pm1.3\pm0.2\pm0.6$ & $\xx7.3\pm1.3\pm0.3\pm0.3$ \\
 $3.0-4.0$ & $  14.9\pm2.4\pm1.3\pm0.6$ & $  12.9\pm1.3\pm0.7\pm0.5$ & $  12.3\pm1.1\pm0.6\pm0.5$ & $\xx8.4\pm0.9\pm1.4\pm0.3$ & $\xx5.9\pm1.1\pm0.3\pm0.2$ \\
 $4.0-5.0$ & $  10.1\pm1.6\pm0.8\pm0.4$ & $\xx9.1\pm0.9\pm0.3\pm0.4$ & $\xx6.7\pm0.8\pm0.4\pm0.3$ & $\xx6.0\pm0.8\pm0.1\pm0.2$ & $\xx3.9\pm0.8\pm0.3\pm0.2$ \\
 $5.0-6.0$ & $\xx6.6\pm1.2\pm0.3\pm0.3$ & $\xx4.1\pm0.6\pm0.3\pm0.2$ & $\xx4.7\pm0.6\pm0.1\pm0.2$ & $\xx3.3\pm0.5\pm0.2\pm0.1$ & $\xx1.4\pm0.5\pm0.2\pm0.1$ \\
 $6.0-7.0$ & $\xx3.9\pm0.8\pm0.4\pm0.2$ & $\xx3.7\pm0.6\pm0.1\pm0.1$ & $\xx2.2\pm0.4\pm0.0\pm0.1$ & $\xx1.3\pm0.3\pm0.1\pm0.1$ & $\xx1.0\pm0.4\pm0.6\pm0.0$ \\
 $7.0-14 $ & $\xx0.9\pm0.1\pm0.0\pm0.0$ & $\xx0.9\pm0.1\pm0.0\pm0.0$ & $\xx0.6\pm0.1\pm0.0\pm0.0$ & $\xx0.4\pm0.1\pm0.0\pm0.0$ & $\xx0.2\pm0.1\pm0.1\pm0.0$ \\
 \bottomrule
  \end{tabular}%
 }\end{small}
\label{tab:JpsiDoubleDiff_event}
\end{sidewaystable}

\begin{table}[tb]
  \caption{ \small
    Nuclear modification factor $R_{\pPb}$ as a function of $y$ with $\pt<14\gevc$.
    The first uncertainty is statistical, 
    the second is the systematic,
    and the third is the uncertainty related to the interpolation error.
  }
 \begin{center}
 \scalebox{1.0}{%
   \begin{tabular}{lcc}
   \toprule
 $R_{\pPb}$ & $-4.0<y<-2.5$ & $2.5<y<4.0$ \\
 \midrule
 Prompt $\jpsi$        & $0.93\pm0.03\pm0.05\pm0.05$ & $0.62\pm0.01\pm0.02\pm0.03$ \\
 $\jpsi$ from $b$      & $0.98\pm0.06\pm0.07\pm0.08$ & $0.83\pm0.02\pm0.04\pm0.07$ \\
 Inclusive $\jpsi$     & $0.93\pm0.03\pm0.05\pm0.05$ & $0.63\pm0.01\pm0.03\pm0.03$ \\
 \bottomrule
 \end{tabular}%
 }\end{center}
\label{tab:RpA}
\end{table}

\begin{table}[tb]
  \caption{ \small 
    Forward-backward production ratio $R_{\mbox{\tiny{FB}}}$ as a function of $|y|$
    with $\pt<14\gevc$.
    The first uncertainty is statistical, 
    the second is the uncorrelated systematic component,
    and the third is the systematic uncertainty correlated between bins.
  }
 \begin{center}
 \scalebox{0.9}{%
   \begin{tabular}{lccc}
   \toprule
 $R_{\mbox{\tiny{FB}}}$ 
                   & $2.5<|y|<3.0$        & $3.0<|y|<3.5$        & $3.5<|y|<4.0$\\
 \midrule
 Prompt $\jpsi$    & $0.72\pm0.05\pm0.05\pm0.04$ & $0.62\pm0.03\pm0.01\pm0.03$ & $0.64\pm0.04\pm0.02\pm0.03$\\ 
 $\jpsi$ from $b$  & $0.92\pm0.11\pm0.08\pm0.05$ & $0.84\pm0.08\pm0.07\pm0.04$ & $0.72\pm0.09\pm0.06\pm0.04$\\ 
 Inclusive $\jpsi$ & $0.74\pm0.05\pm0.06\pm0.04$ & $0.64\pm0.03\pm0.01\pm0.03$ & $0.65\pm0.04\pm0.01\pm0.03$\\ 
 \bottomrule
 \end{tabular}%
 }\end{center}
\label{tab:RFB_y}
\end{table}
\begin{table}[tb]
  \caption{ \small 
    Forward-backward production ratio $R_{\mbox{\tiny{FB}}}$ 
    as a function of $p_\mathrm{T}$ with $2.5<|y|<4.0$.
    The first uncertainty is statistical, 
    the second is the uncorrelated systematic component,
    and the third is the systematic uncertainty correlated between bins.  
  }
 \begin{center}
 \scalebox{0.9}{%
   \begin{tabular}{lccc}
   \toprule
   \pt [\gevc] & $R_{\mbox{\tiny{FB}}}$ (prompt \jpsi) & $R_{\mbox{\tiny{FB}}}$ (\jpsi from $b$) & $R_{\mbox{\tiny{FB}}}$ (inclusive \jpsi) \\
   \midrule
 $0.0-1.0$& $0.62\pm0.06\pm0.02\pm0.03$ & $0.84 \pm 0.18 \pm0.05\pm0.04$ & $0.63 \pm 0.06\pm0.02\pm0.03$ \\
 $1.0-2.0$& $0.58\pm0.04\pm0.02\pm0.03$ & $0.97 \pm 0.15 \pm0.15\pm0.05$ & $0.61 \pm 0.04\pm0.03\pm0.03$ \\
 $2.0-3.0$& $0.60\pm0.04\pm0.02\pm0.03$ & $0.85 \pm 0.12 \pm0.14\pm0.04$ & $0.62 \pm 0.04\pm0.02\pm0.03$ \\
 $3.0-4.0$& $0.66\pm0.04\pm0.01\pm0.03$ & $0.96 \pm 0.14 \pm0.11\pm0.05$ & $0.69 \pm 0.04\pm0.02\pm0.03$ \\
 $4.0-5.0$& $0.75\pm0.06\pm0.02\pm0.04$ & $0.89 \pm 0.16 \pm0.12\pm0.04$ & $0.76 \pm 0.05\pm0.02\pm0.04$ \\
 $5.0-6.0$& $0.75\pm0.07\pm0.02\pm0.04$ & $0.77 \pm 0.16 \pm0.04\pm0.04$ & $0.75 \pm 0.07\pm0.03\pm0.04$ \\
 $6.0-7.0$& $0.77\pm0.08\pm0.02\pm0.04$ & $0.91 \pm 0.22 \pm0.14\pm0.05$ & $0.78 \pm 0.09\pm0.03\pm0.04$ \\
 $7.0-8.5$& $0.83\pm0.11\pm0.03\pm0.04$ & $1.08 \pm 0.27 \pm0.09\pm0.05$ & $0.87 \pm 0.11\pm0.03\pm0.04$ \\
 $8.5-14 $& $0.67\pm0.10\pm0.03\pm0.03$ & $0.74 \pm 0.19 \pm0.04\pm0.04$ & $0.68 \pm 0.09\pm0.03\pm0.03$ \\
 \bottomrule
 \end{tabular}%
 }\end{center}
\label{tab:RFB_pt}
\end{table}
\clearpage